\documentclass[twocolumn,showpacs,amsmath,nofootinbib,pra,aps,amssymb,longbibliography]{revtex4-2}  
\usepackage[T1]{fontenc}
\usepackage[utf8]{inputenc}
\usepackage{times}
\usepackage{color} 
\usepackage{array}
\usepackage{amssymb,amsmath}
\usepackage{amsfonts}
\usepackage{amsbsy}
\usepackage{wasysym}
\usepackage[pdftex]{graphicx} 
\usepackage{bm}
\usepackage{float}
\usepackage{dcolumn}
\usepackage{siunitx}
\usepackage[caption=false]{subfig}
\usepackage{mhchem}

\graphicspath{{images/}}
\usepackage[unicode,breaklinks]{hyperref}
\hypersetup{
    unicode=true,
    plainpages=false, 
    colorlinks=true,
    linkcolor=blue,
    citecolor=blue,
    filecolor=black,
    urlcolor=blue
}
\newcommand\D{\mathop{}\!d}
\newcommand\ba{\mathbf{a}}

\newcommand\bx{\mathbf{x}}
\newcommand\br{\mathbf{r}}
\newcommand\bv{\mathbf{v}}
\newcommand\brho{\boldsymbol{\rho}}

\newcommand\ME{\mathcal{E}}
\newcommand\MF{\mathcal{F}}
\newcommand\MC{\mathcal{C}}
\newcommand\uc{\mathrm{uc}}

\newcommand\gammaQF{\gamma_{\mathrm{QF}}}

\newcommand\gQF{g_{\mathrm{QF}}}

\newcommand\gdd{g_\mathrm{dd}}
\newcommand\add{a_\mathrm{dd}}

\newcommand\gs{g_s} % not roman
\newcommand\as{a_s} % not roman
\newcommand\Hsp{H_\mathrm{sp}}
\newcommand\intuc{\int_\uc\!\!}

\newcommand\psiCM{\psi_\mathrm{CM}}
\newcommand\psiDA{\psi_\mathrm{DA}}
\DeclareMathOperator{\erfc}{erfc}

\newcommand\UDt{\tilde{U}_\sigma}
\begin{document}
\title{Two-dimensional supersolidity in a planar dipolar Bose gas} 
\author{B. T. E. Ripley, D. Baillie, and P. B. Blakie}
	
	\affiliation{Dodd-Walls Centre for Photonic and Quantum Technologies, New Zealand}
	\affiliation{Department of Physics, University of Otago, Dunedin 9016, New Zealand}
 
\date{\today}
\begin{abstract}
We investigate the crystalline stationary states of a dipolar Bose-Einstein condensate in a planar trapping geometry. Our focus is on the ground state phase diagram in the thermodynamic limit, where triangular, honeycomb and stripe phases occur. 
We quantify the superfluid fraction by calculating the non-classical translational inertia, which allows us to identify favorable parameter regimes for observing supersolid ground states. 
We develop two simplified theories to approximately describe the ground states, and consider the relationship to roton softening in the uniform ground state. This also allows us to extend the phase diagram to the low density regime. While the triangular and honeycomb states have an isotropic superfluid response tensor, the stripe state exhibits anisotropic superfluidity.
\end{abstract}

\maketitle

\section{Introduction}
Experiments with dipolar Bose-Einstein condensates (BECs) have produced supersolid states \cite{Tanzi2019a,Bottcher2019a,Chomaz2019a} which exhibit superfluidity and crystalline order. Most experimental work has used elongated (cigar-shaped) potentials where one-dimensional (1D) modulation of the spatial density occurs. However, recent work has seen the production of a two-dimensional (2D) supersolid state \cite{Norcia2021a}. By varying the geometry of the harmonic trap, experiments have been able to explore a structural phase transition between a long 1D and 2D supersolid. 

The supersolid transition in these systems arises from an interplay between the short-ranged contact interactions, long-ranged dipole-dipole interactions (DDIs), and quantum fluctuations \cite{Chomaz2023a}. For the 1D case, the phase diagram and nature of the transition has been explored in a number of works \cite{Roccuzzo2019a,Blakie2020b,Biagioni2022a,Ilg2023a,Smith2023a}. In such a system the transition is found to be continuous over a broad intermediate range of densities, and coincides with the softening of a roton excitation in the uniform BEC state (cf.~\cite{Sepulveda2008a}). 
The continuous transition makes it feasible to dynamically produce the supersolid state starting from the BEC by ramping the $s$-wave scattering length across the transition using a Feshbach resonance.

Two-dimensional supersolids will offer a number of interesting new features, such as different crystal structures \cite{Lu2015a,Zhang2019a}, a richer excitation spectrum with three gapless excitation branches \cite{Watanabe2012a}, and the possibility for the supersolid to host vortices \cite{Klaus2022a}. In Ref.~\cite{Zhang2019a}, Zhang \textit{et al.}~produced a phase diagram for a dipolar Bose gas in a planar system in the thermodynamic limit. This work showed that a triangular supersolid state occurs in the low density regime and a honeycomb supersolid occurs at higher densities. The transition from the uniform superfluid to these states is discontinuous except at an intermediate density where the states coexist at a critical point. These results already distinguished the dipolar case from earlier work on a 2D system with soft-core interactions, where a phase diagram emerges with only uniform superfluid and triangular supersolid ground states and a discontinuous transition between these states \cite{Pomeau1994a,Saccani2012a,Hsueh2012a,Kunimi2012a,Macri2013a,Prestipino2018a}. A number of groups have studied the phase diagram in the finite system with a three-dimensional pancake-shaped harmonic trap \cite{Baillie2018a,Zhang2021a,Hertkorn2021a,Norcia2021a,Poli2021a,Schmidt2022a} and have found phase diagrams (see \cite{Zhang2021a,Hertkorn2021a,Schmidt2022a}) with ground states exhibiting triangular, stripe (or labyrinth), and honeycomb patterns. Triangular droplet array ground states should be accessible to current experiments with magnetic atoms, with the other phases requiring larger atom numbers or stronger DDIs. These other phases may be easier to explore with molecular gases \cite{Schmidt2022a}. The Feshbach ramps used to prepare 1D dipolar supersolids will generally not work as well for the 2D case due to the (generally) discontinuous superfluid-supersolid transition, in which a metastable excited state is produced by the nucleation of small incoherent droplets \cite{Kadau2016a,Bisset2015a,Xi2016a,Blakie2016a,Ferrier-Barbut2018a}. Directly cooling into the supersolid state has been proposed as an alternative method to avoid this issue \cite{Bland2022a}.

\begin{figure}[htbp]
	\centering
	\includegraphics[width=\linewidth]{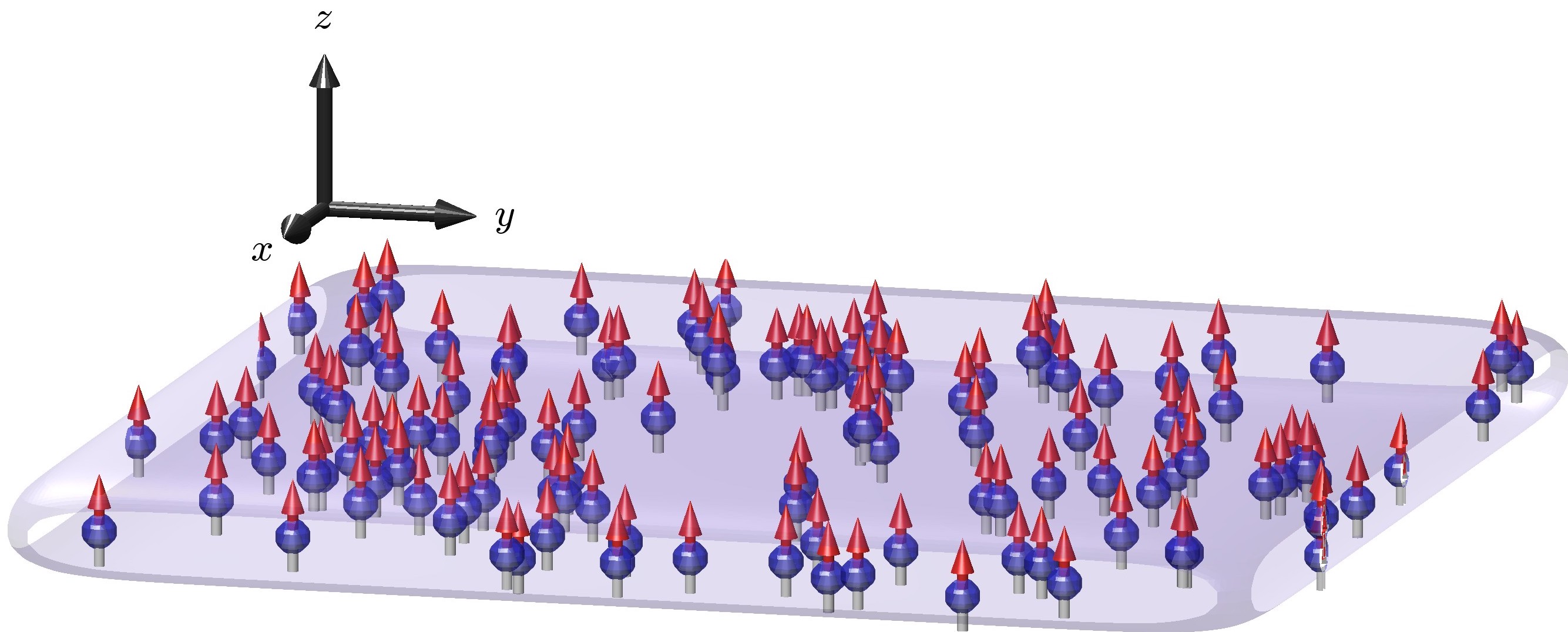}
    \caption{Schematic figure of the planar dipolar Bose gas with the atoms harmonically confined along $z$ (parallel to the dipole polarization direction) and free to move in the $xy$-plane.}
	\label{f:schematic}
\end{figure}

In this paper we consider the stationary states of a dipolar Bose gas in a uniform planar potential (see Fig.~\ref{f:schematic}). This system is the thermodynamic limit of the pancake shaped traps and was originally considered in Ref.~\cite{Zhang2019a}. Notably, we locate a region where a stripe phase is the ground state. This was not found in the earlier work, and provides a connection to the stripe or labyrinth states found in trapped studies. Our main results are based on theory using a variational Gaussian treatment of the axially confined direction. The planar direction is treated numerically with the unit cell geometry constrained to select states with triangular (including honeycomb) or stripe geometry. We also develop analytic approximations suitable for describing the stripe state and a triangular droplet crystal. We introduce a rigorous method to calculate the superfluidity of these states by analysing their nonclassical translational inertia. The resulting superfluid fraction is a rank-2 tensor which is isotropic for the triangular and honeycomb states but anisotropic for the stripe case.

\section{Theory}\label{Sec:Theory}
\subsection{System}
We consider a magnetic dipolar condensate free in the radial plane and harmonically confined with a trap frequency $\omega_z$ in the $z$ direction, described by the single particle Hamiltonian
\begin{equation}
	\Hsp = -\frac{\hbar^2\nabla^2}{2m} + \frac12 m\omega_z^2 z^2.
\end{equation}
The atoms have an interaction of the form
\begin{equation}
	U(\br) = g_s\delta(\br) + \frac{3\gdd}{4\pi r^3}(1-3\cos^2\theta),
\end{equation}
where the short-ranged interaction is governed by the coupling constant $\gs = 4\pi \as \hbar^2/m$ with $\as$ being the $s$-wave scattering length, and the DDI by $\gdd = \mu_0 \mu_m^2/3$ with $\mu_m$ being the magnetic moment. We take the magnetic dipoles to be polarized along $z$ by an external bias field, and $\theta$ is the angle between $\br$ and the $z$ axis. 

The mean-field energy functional describing the system is
\begin{align}\label{e:efunc0}
    E &= \int\D\bx\,\Psi^*\biggl[\Hsp + \frac12\Phi(\bx) + \frac25\gammaQF|\Psi|^3\biggr]\Psi,
\end{align}
where 
\begin{align}
	\Phi(\bx) &= \int \D\bx'\, U(\bx-\bx')|\Psi(\bx')|^2.
\end{align}
Here we also account for the effects of quantum fluctuations in the term with coefficient \cite{Lima2011a,Lima2012a,Ferrier-Barbut2016a,Wachtler2016a,Bisset2016a,Saito2016a}
\begin{equation} 
	\gammaQF = \frac{128\sqrt{\pi}\hbar^2\as^{5/2}}{3m}\left[1+\frac32\left(\frac{\add}{\as}\right)^2\right].
\end{equation}

\subsection{Unit cell treatment}\label{Sec:UCtreatment}
Our interest is in minimizing the energy per particle for a system with fixed areal density $n$. 
We find that such solutions are either uniform or have a periodic modulated (crystalline) density in the plane. In particular, we consider modulated states both on a triangular unit cell, allowing for a 2D crystal structure, and on a linear unit cell, allowing for a 1D stripe crystal.
 
We take the field to be of the general form
\begin{align}
\Psi(\bx) = \sqrt{n}\psi(\brho)\chi_\sigma(z),
\end{align}
where we employ a variational Gaussian treatment of the axial field
\begin{align}
\chi_\sigma(z)=\frac{1}{\pi^{1/4} \sigma^{1/2}}e^{-z^2/2\sigma^2},
\end{align}
with variational width $\sigma$. The planar field $\psi(\brho)$, with $\brho=(x,y)$, is treated in various ways subject to the nature of the unit cell, as discussed below. The energy per particle is given by $\ME = \ME_z + \ME_\perp$, with axial component $\ME_z=\hbar\omega_z({a_z^2}/{4\sigma^2} + {\sigma^2}/{4a_z^2})$ where $a_z=\sqrt{\hbar/m\omega_z}$, and planar component
\begin{align}\label{e:efunctr}
    \ME_\perp &= \intuc \frac{\D\brho}A \,\psi^*\biggl[ -\frac{\hbar^2\nabla^2_\rho}{2m} + \frac12\Phi_\sigma(\brho) + \frac25\gQF n^{3/2}|\psi|^3\biggr]\psi.
\end{align}
This expression for $\ME$ is obtained by integrating out the axial degrees of freedom from Eq.~(\ref{e:efunc0}), and we have introduced $\gQF = \gamma_5\gammaQF$, $\gamma_5=\sqrt{2/5}/(\sqrt\pi\sigma)^{3/2}$,
\begin{align}
\Phi_\sigma(\brho) = n\MF_\rho^{-1}\{\UDt(k_\rho)\MF_\rho\{|\psi(\brho)|^2\}\},
\end{align}
 where $\MF_\rho$ is the 2D Fourier transform, and
\begin{align}
    \UDt(k_\rho) &= \{g_s + \gdd[2 - 3G_0(\sigma k_\rho/\sqrt2)] \}\gamma_4,\label{e:UDTkrho}
\end{align}
with $G_0(q)=\sqrt\pi qe^{q^2}\erfc(q)$, and $\gamma_4=1/\sqrt{2\pi}\sigma$.
In Eq.~(\ref{e:efunctr}) we carry out the integration over a unit cell, with $A$ being the unit cell area. 
Stationary solutions of Eq.~(\ref{e:efunctr}) satisfy the time-independent extended Gross-Pitaevskii equation (eGPE)
\begin{align}
    \mu\psi &= \mathcal{L}_{\mathrm{GP}}\psi,\label{tiGPE}
\end{align}
where $\mu$ is the chemical potential and we have introduced the eGPE operator
\begin{align} 
    \mathcal{L}_{\mathrm{GP}}= -\frac{\hbar^2\nabla^2_\rho}{2m} + \Phi_\sigma(\brho) + \gQF n^{3/2}|\psi|^3 .
\end{align}

\subsubsection{Uniform state}
The uniform state has $\psi(\brho)=1$. In this case the unit cell is arbitrary and the planar energy reduces to
\begin{equation}
    \ME_\perp = \frac12(\gs+2\gdd)\gamma_4 n + \frac25\gQF n^{3/2}. 
\end{equation}
Its properties are thus determined by minimizing $\ME$ for fixed $n$ against the single parameter $\sigma$.

\subsubsection{Triangular unit cell}
Here we take $\psi(\brho)$ periodic on a unit cell defined by the two direct lattice vectors $\ba_1$ and $\ba_2$, with $|\ba_1|=|\ba_2|\equiv a$, angle $\pi/3$ between $\ba_1$ and $\ba_2$, unit cell area $A=|\ba_1\times \ba_2|=\sqrt3 a^2/2$. The planar field has the normalization condition $\intuc \D\brho\, |\psi|^2 =A$, where the integration is taken over the unit cell. We note that the triangular unit cell can support both triangular and honeycomb states (see Sec.~\ref{s:results}).

The planar energy per particle is evaluated numerically from Eq.~(\ref{e:efunctr}), and stationary states are determined by finding local minima of $\ME$ for fixed $n$, with respect to $\psi$, $\sigma$ and $a$.

\subsubsection{Stripe state}
The striped state is uniform in one direction, which we take to be $y$, and we thus set $\psi(\brho)\to\psi(x)$, i.e.,~being periodic on a 1D unit cell in the $x$ direction with length $L$, $\uc=\{-\tfrac12L \le x < \tfrac12 L\}$. The normalization is $\intuc \D x \,|\psi(x)|^2 = L$.
The planar energy per particle is 
\begin{equation}
    \ME_\perp = \intuc \frac{\D x}L \,\psi^*\biggl[-\frac{\hbar^2}{2m}\frac{\D^2}{\D x^2} + 
\frac12 \Phi_\sigma(x) + \frac25\gQF|\psi|^3n^{3/2}\biggr]\psi,
\end{equation}
with $\Phi_\sigma(x) = n\MF_x^{-1}\{\UDt(k_x)\MF_x\{|\psi(x)|^2\}\}$, where $\MF_x$ is the 1D Fourier transform.
As for the triangular case, we determine stationary states by locating minima of $\ME$ for fixed $n$, with respect to $\psi$, $\sigma$ and $L$.
\subsection{Numerical method}
In our numerics we represent the planar field in terms of plane waves associated with the reciprocal lattice vectors of the unit cell. This allows us to evaluate the kinetic energy and Fourier transforms (for the interaction terms) with spectral accuracy. 
We mainly use a gradient flow method (e.g.,~see Refs.~\cite{Bao2004a,Lee2021a,Smith2023a}) to determine $\psi$ and minimize the energy per particle. During the flow we also optimize $\sigma$ and the unit cell size.
For cases where the convergence is slow we employ a Newton-Krylov method between applications of gradient flow.
We observe that states on the triangular unit cell exhibit 6-fold mirror symmetry (such that a single dodecant of the Wigner-Seitz unit cell sufficiently characterizes the state), and enforce this symmetry to follow metastable states.

\subsection{Superfluid tensor}
Due to the spontaneous breaking of translational invariance, the modulated ground state can have a reduced superfluid fraction, even at zero temperature (e.g.,~see \cite{Leggett1970a,Leggett1998a}). We can quantify the superfluidity through the non-classical translational inertia, i.e.,~the momentum response of the system to moving walls. 
For walls moving sufficiently slowly at velocity $\mathbf{v}=(v_x,v_y)$, the normal fluid will move with the walls while the superfluid will remain at rest. 
 The superfluid fraction is related to the reduction in translational inertia from the classical value $M=Nm$, where $N$ is the number of atoms in the sample. We can define the superfluid fraction tensor as 
 \begin{align}\label{e:SF}
	f_{s,ij} =\delta_{ij} - \lim_{v \to 0} \frac{1}{M}\frac{ \partial {P}_{i} }{\partial v_j},\qquad i,j\in\{x,y\}
\end{align}
where $v=|\bv|$, and $\mathbf{P}=-i(n\hbar/2)\int d\bm{\rho}\,(\psi^*\bm{\nabla}_\rho\psi-\psi\bm{\nabla}_\rho\psi^*)$ is the planar momentum (cf.~\cite{Ancilotto2013a,Roccuzzo2019a}).
Within our formalism we can calculate the equilibrium state in a moving frame by finding stationary states of the eGPE
\begin{align}
\mu_{\mathbf{v}}\psi_{\mathbf{v}} = 
\mathcal{L}_{\mathrm{GP}}\psi_{\mathbf{v}} +
i\hbar\mathbf{v}\cdot\bm{\nabla}_\rho\psi_{\mathbf{v}}.
\end{align}
For small velocities we obtain the first order perturbative result $\psi_{\mathbf{v}} =\psi_0e^{i\phi_1}$, where $\psi_0$ is the ground state in the absence of motion and $\phi_1$ is determined by
\begin{align}
\frac{\hbar}{m}\bm{\nabla}_\rho\cdot\left(|\psi_0|^2\bm{\nabla}_\rho\phi_1\right)= {\mathbf{v}}\cdot\bm{\nabla}_\rho|\psi_0|^2.\label{phi1}
\end{align}
Here the functions $\psi_0$ and $\phi_1$ are periodic on the unit cell and (with the freedom of a global phase choice) we take $\phi_1$ to have an average value of zero in the unit cell.
This linear system can be solved for the phase correction \cite{Saslow1976a,Blakie2023b} and the superfluid tensor determined from Eq.~(\ref{e:SF}) using $\mathbf{P}=n\hbar\int\,d\bm{\rho}\,|\psi_0|^2\bm{\nabla}_\rho\phi_1$. In practice, because the functions appearing in this expression for $\mathbf{P}$ are all periodic, we can evaluate this result on a single unit cell. This allows us to employ the same spectral method we use to obtain $\psi$ to calculate $\phi_1$ by solving the linear system Eq.~(\ref{phi1}).

\section{Results}\label{s:results}
\subsection{Phase diagram}\label{sec:pd}
\begin{figure}[htbp]
	\centering
	\includegraphics[width=\linewidth]{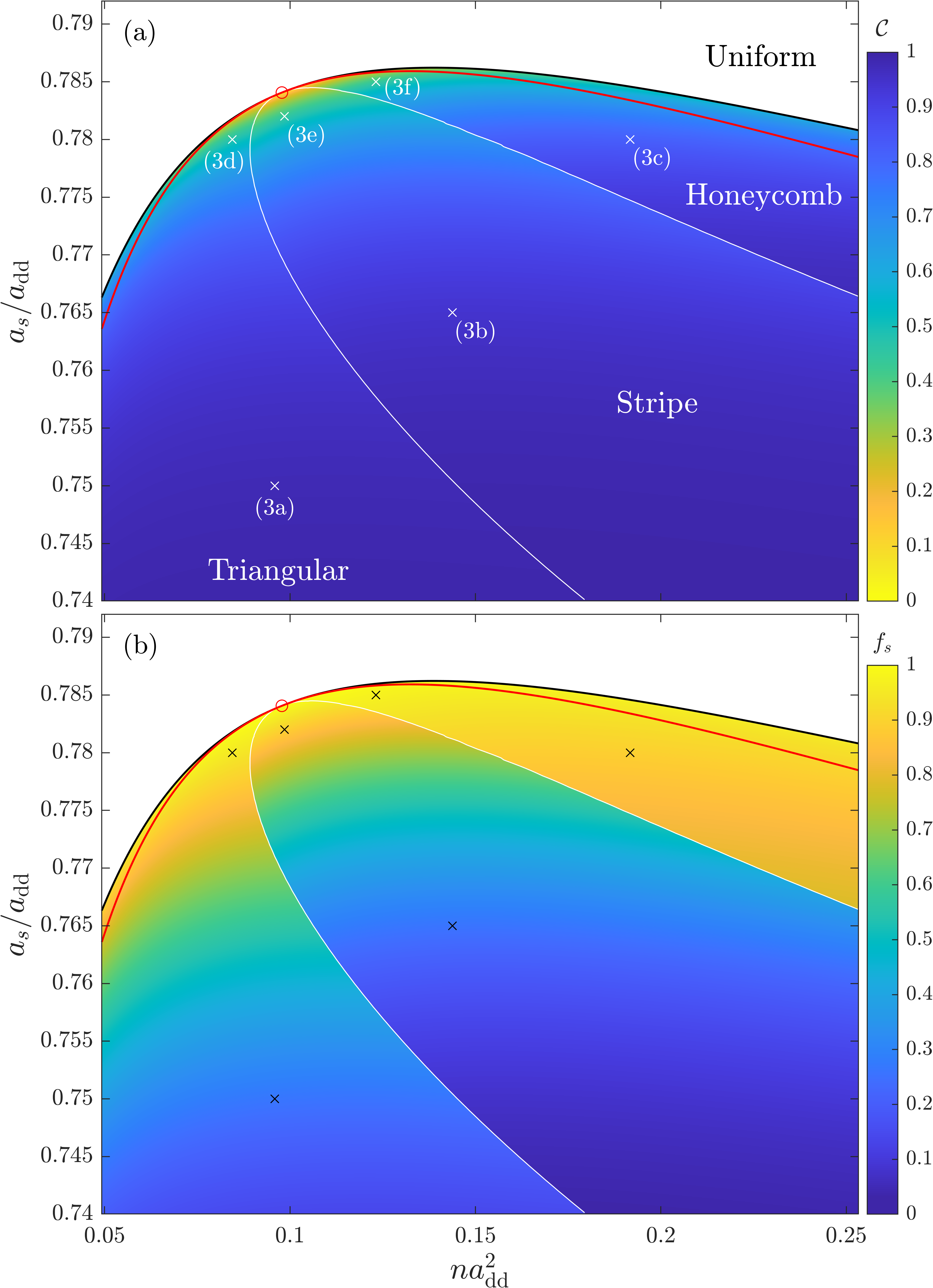}
    \caption{Ground state phase diagrams depicting (a) the contrast $\MC$ and (b) the superfluid fraction $f_s$. Discontinuities reveal three different ground state unit cell geometries, which we denote triangular, stripe, and honeycomb (example states marked $\times$ are shown in Fig.~\ref{f:n3}). Above the black melting line, the uniform unmodulated state has the lowest energy (note that in this phase, $\MC=0$ and $f_s=1$). The roton boundary (red line) meets the uniform state only at the critical point (red circle). Results for \ce{^{164}Dy} using $\add=130.8 a_0$, with $\omega_z/2\pi = 72.4\:\unit{\hertz}$.}
	\label{f:pd}
\end{figure}

\begin{figure}[htbp]
	\centering
    \includegraphics[trim=182 462 188 125,clip=true,width=\linewidth]{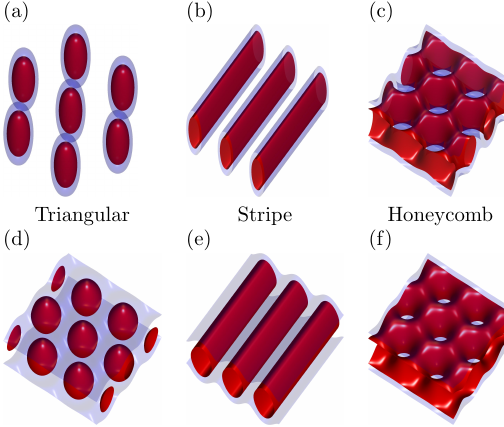} 
	\caption{Selected modulated ground-states as indicated on the phase diagrams in Fig.~\ref{f:pd} with $\times$. Isosurfaces at 50\% (red) and 25\% (blue) of the peak density. Parameters for the ground-states are
		(a) $(n\add^2,\as/\add) \approx (0.0959,0.750)$, 
		(b) $(0.144,0.765)$, 
		(c) $(0.192,0.780)$,
		(d) $(0.0844,0.780)$, 
		(e) $(0.0985,0.782)$, and 
        (f) $(0.123,0.785)$.}
	\label{f:n3}
\end{figure}

A phase diagram showing the ground-state phase as a function of density and $s$-wave scattering length for a fixed trap frequency $\omega_z$ is shown in Fig~\ref{f:pd}(a).
The shading color indicates the density contrast to characterize the density modulation of the ground-states in the $xy$-plane. This is defined as
\begin{equation}
	\MC = \frac{\max{|\psi|^2}-\min{|\psi|^2}}{\max{|\psi|^2}+\min{|\psi|^2}},
\end{equation}
so that a uniform state would yield a contrast of $\MC=0$, a state for which the planar density $n|\psi|^2$ somewhere goes to zero yields $\MC=1$, and a crystalline state has $\MC>0$.
Using the different unit cell treatments we assess which of the uniform or various geometry modulated states have the lowest energy.
At high scattering length the ground state is a uniform (unmodulated) BEC, but as the scattering length decreases, a melting line is crossed and the system transitions to one of three modulated states, examples of which are shown in Fig~\ref{f:n3}.
In addition to the stripe state [Fig~\ref{f:n3}(b)], the ground state in the triangular cell can be a triangular array of (droplet-like) density maxima [Fig~\ref{f:n3}(a)] or a honeycomb configuration [Fig~\ref{f:n3}(c)]. The honeycomb state is often similar to an inversion of the density of the triangular state.
The work of Ref.~\cite{Zhang2019a} found triangular and honeycomb phases for the system we consider here.

\subsection{Superfluidity}
In Fig.~\ref{f:pd}(b) we present results for the superfluid fraction across the phase diagram. 
The superfluid fraction is unity in the uniform phase and remains high in the modulated phases sufficiently close to the melting line. In the triangular and honeycomb phases, the superfluid fraction tensor is isotropic, such that $f_{s,ij}=f_s\delta_{ij}$ (e.g.,~see \cite{Sepulveda2010a,Blakie2023b}). 
In the stripe phase, the superfluid fraction can be calculated analytically: from Eq.~(\ref{phi1}) we obtain
\begin{align}
\frac{\hbar}{m}\phi_1^\prime(x)=v_x\left(1-\frac{c_s}{|\psi_0(x)|^2}\right),
\end{align}
where the integration constant $c_s=L[\intuc \,dx/|\psi_0(x)|^2]^{-1}$ is determined by the periodicity requirement for $\phi_1$.%
\footnote{Note that this constant has a form analogous to those of the Leggett bounds, shown as Eqs.~(A5) \& (A7) of Ref.~\cite{Leggett1998a}, which are exact measures of the superfluidity in 1D or when the wave function is separable \cite{Leggett1998a,Sepulveda2008a}.}
Evaluating Eq.~(\ref{e:SF}) with this result, we obtain that the superfluid fraction tensor is diagonal with
\begin{align}
f_{s,xx}=c_s,\quad
f_{s,yy}=1. 
\end{align}
For nonzero contrast, $c_s<1$ and the superfluid tensor is anisotropic, reflecting a difference in the tendency of the system to resist motion parallel and perpendicular to the stripes. 

For the results in Fig.~\ref{f:pd}(b) we use the smallest superfluid fraction for the stripe state (i.e.,~taking $f_s=f_{s,xx}$). 
 We observe that generally the superfluid fractions of the triangular and honeycomb phases are higher relative to $f_{s,xx}$ for the stripe phase. It is interesting to contrast these results against the 2D soft-core system: there, the triangular state is the only modulated phase, with stripe and honeycomb states always being metastable excited states \cite{Prestipino2018a}\footnote{A finite moving frame velocity can favor a stripe ground state \cite{Kunimi2012a}.}. This serves to emphasize the richness of the planar dipolar system. Additionally, the metastable stripe state in the soft-core system tends to have a higher superfluid fraction (along the modulated direction) than the triangular state for the same parameters \cite{Blakie2023b}.
 
\begin{figure}[htb!]
	\centering
	\includegraphics[width=0.95\linewidth]{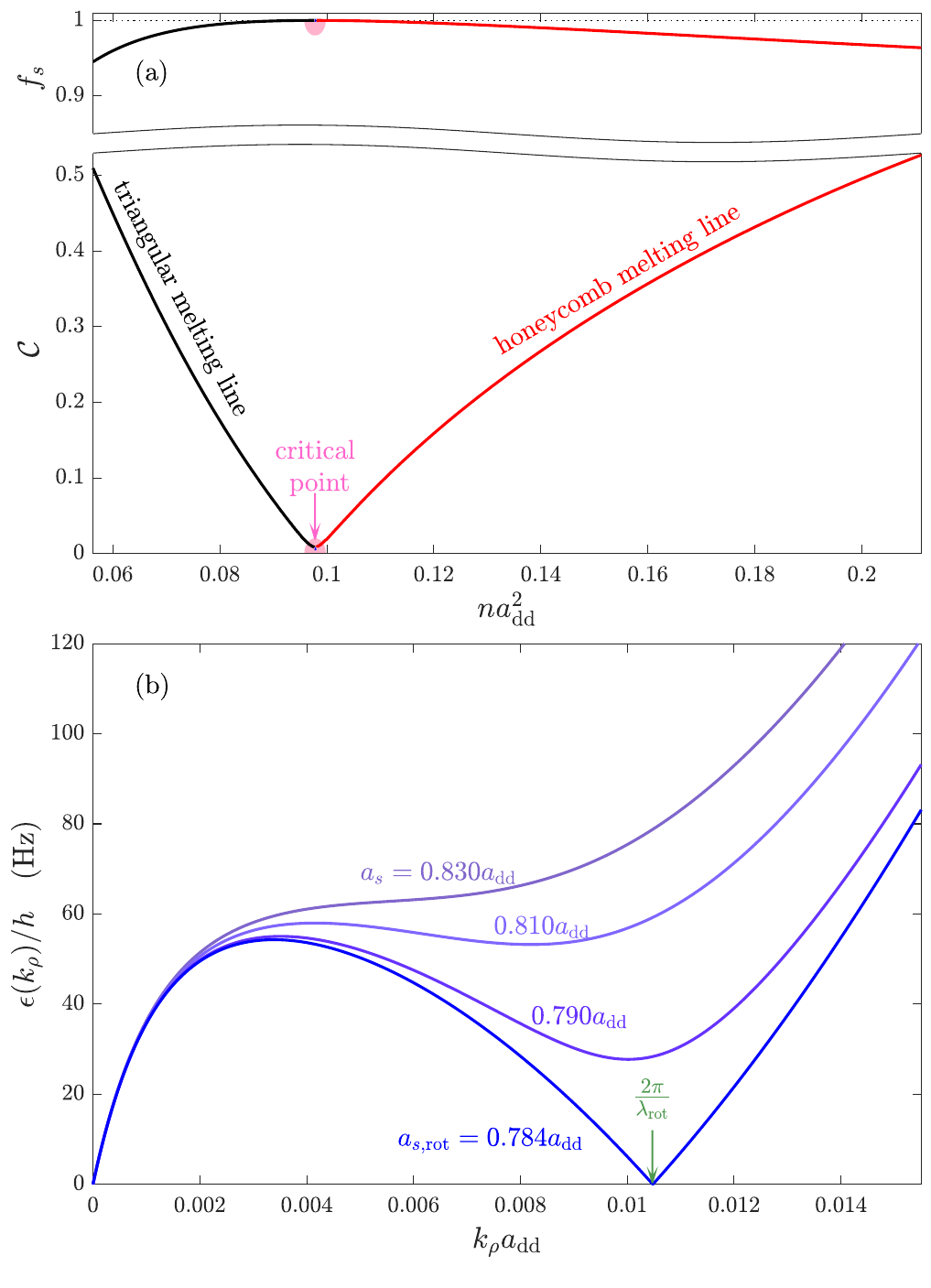} 
	\caption{(a) Contrast and superfluidity of the modulated states along the melting line. (b) Dispersion relation for the uniform condensate at the critical density, $n\add^2 = 0.0978$. Local minima represent roton excitations, the energy of which vanish as $\as \rightarrow a_{s,\mathrm{rot}}=0.784\add$. 
	Other parameters for the calculations as specified in Fig.~\ref{f:pd}.}
	\label{f:crplot}
\end{figure}

\subsection{Critical point}

In the phase diagram we find that the different ground states are separated by first-order phase transitions, with a continuous critical point where all four phases meet. 
Our physical parameters for this phase diagram are the same as those used in Ref.~\cite{Zhang2019a}, and we find the critical point at $(n_\mathrm{crit}\add^2,a_{s,\mathrm{crit}}/a_{\mathrm{dd}})\approx(0.0978,0.784)$.
This is slightly lower in density and scattering length than the value of $(0.110,0.793)$ found in \cite{Zhang2019a}, and is likely due to our variational treatment of the axial direction. In Sec.~\ref{sec:TF} we also consider a Thomas-Fermi approximation for the axial direction and identify the critical point within that approximation.

In Fig.~\ref{f:crplot}(a) we present results for the contrast and superfluid density along the melting line [i.e.,~along the black line in Figs.~\ref{f:pd}(a) and (b)] over a density region containing the critical point. 
Approaching the critical point the contrast vanishes and the superfluid fraction approaches unity, such that the transition to the uniform state is continuous at this point.

\subsection{Roton softening}\label{Sec:roton}
The quasi-particle excitations of a ground state described by the eGPE can be obtained by solving the Bogoliubov-de Gennes equations (e.g.,~see \cite{Baillie2017a}). For the uniform BEC state the excitations are plane waves and with a dispersion relation 
\begin{align}
\epsilon(k_\rho) = \sqrt{\frac{\hbar^2k_\rho^2}{2m}\Delta(k_\rho)},\label{edispersion}
\end{align}
where
\begin{align}
    \Delta(k_\rho) =\frac{\hbar^2k_\rho^2}{2m}+ 2n\tilde{U}_\sigma(k_\rho)+3\gQF n^{3/2}.
    \label{e:Delta}
\end{align}
This is an extension of the results of \cite{Baillie2015b} to include quantum fluctuations (cf.~Ref.~\cite{Blakie2020b}).
The dipole-dominant regime occurs when $\as<\add$, and in this regime the uniform condensate state can exhibit a roton-like excitation which manifests as a local minimum in the excitation spectrum at a finite non-zero wave vector, as shown in Fig.~\ref{f:crplot}(b). By decreasing $\as$ further, the roton softens to zero energy and we can identify a critical value $a_{s,\mathrm{rot}}(n)$ with a corresponding roton wavelength $\lambda_\mathrm{rot}(n)$ of the soft excitation. Note that for $\as<a_{s,\mathrm{rot}}(n)$ the uniform state is dynamically unstable. 
The roton instability line is indicated on the phase diagram in Fig.~\ref{f:pd}, and we observe that the roton line generally lies below the melting line, with the two lines touching at the critical point. 
The results in Fig.~\ref{f:crplot}(b) are at the critical density, and confirm that the roton softening coincides with the critical point. This scenario is similar to the continuous transition in 1D supersolids \cite{Sepulveda2008a,Roccuzzo2019a, Blakie2020b}, although in that case the continuous transition holds for a range of densities.

\begin{figure}
	\centering
	\includegraphics[width=\linewidth]{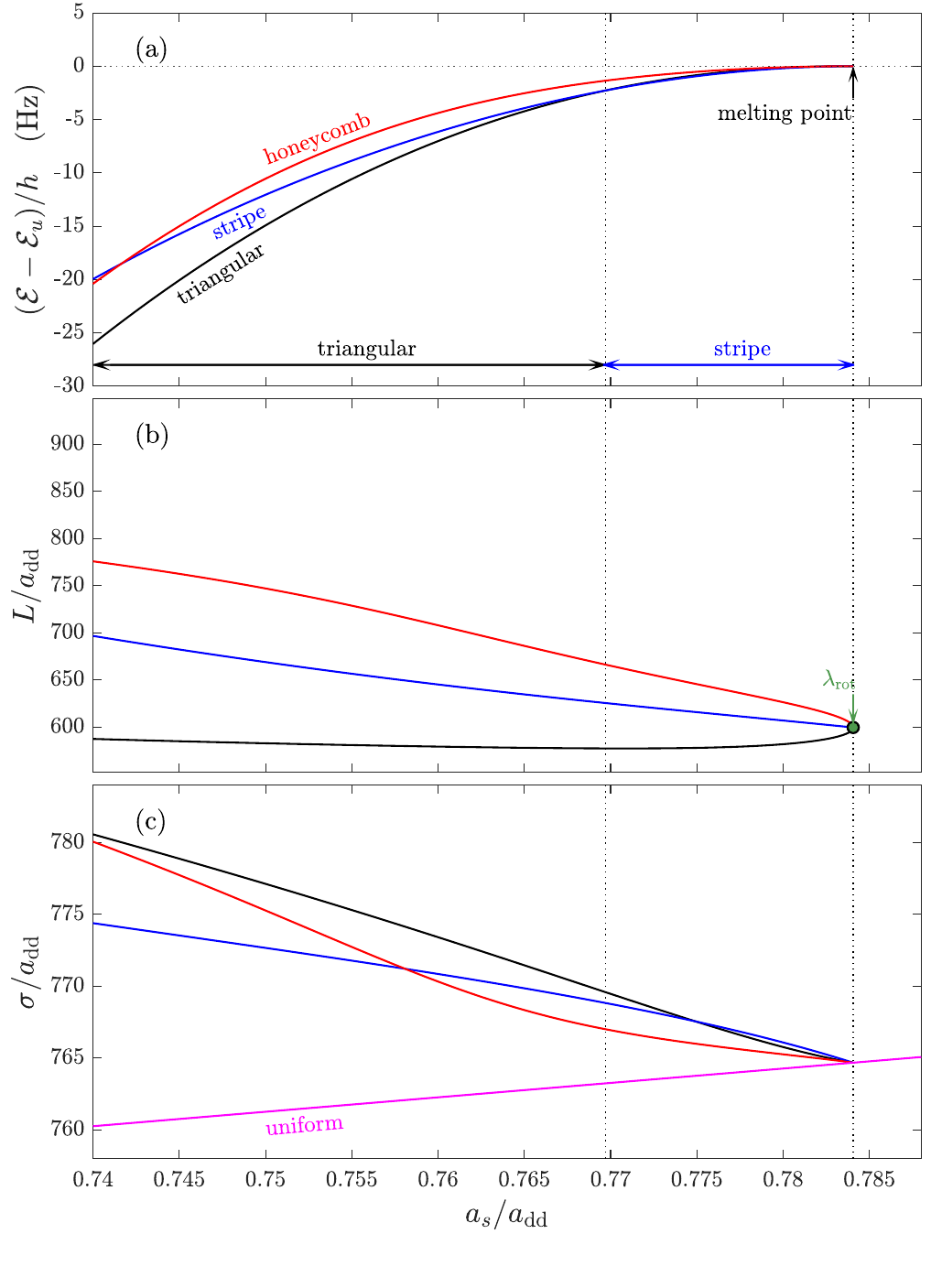}
	\caption{Properties of the various constrained geometry solutions for a system at the critical density, $n\add^2 = 0.0978$. 
		(a) The energy per particle relative to the uniform BEC state. 
		(b) The unit cell length scale $L$, showing that these converge to the roton wavelength at the critical point. 
		(c) The variational Gaussian width parameter.
		Other parameters for the calculations as specified in Fig.~\ref{f:pd}.}	
		
	\label{f:propncrit}
\end{figure}

\begin{figure}
	\centering
	\includegraphics[width=\linewidth]{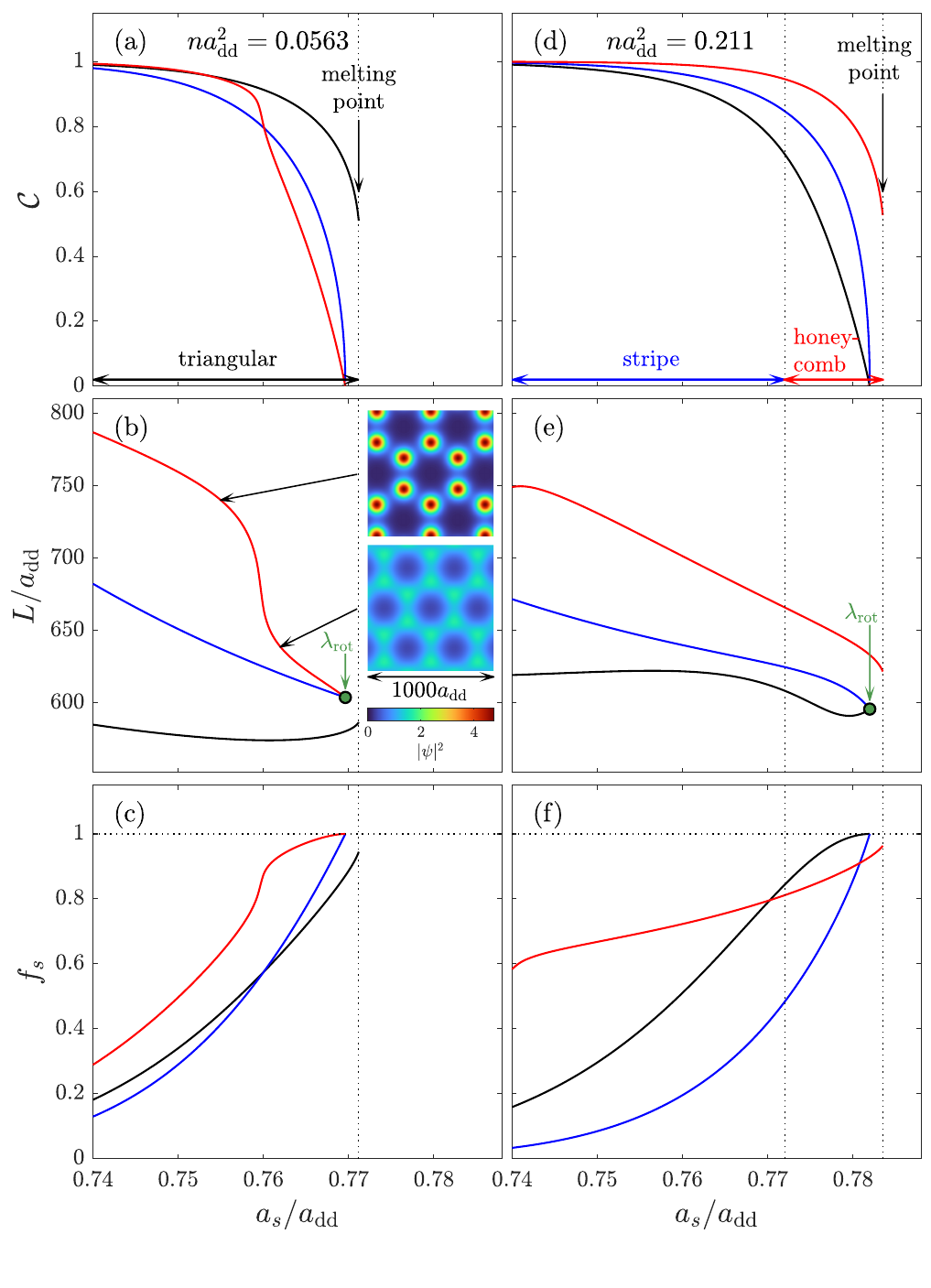}
	\caption{Contrast, unit cell size, and superfluid fraction below (a-c) and above (d-f) the critical density. The insets in (b) show that the honeycomb state forms separated droplets at high contrast (top), and interconnected droplets with smaller cell size at lower contrast (bottom).}
	\label{f:multiplot}
\end{figure}

\subsection{Solution properties}\label{Sec:solnprops}

Here we consider some general properties of the ground and metastable states across the phase diagram.

The results in Fig.~\ref{f:propncrit} are at the critical density.
The energy per particle [Fig.~\ref{f:propncrit}(a)] reveals the transition in ground state from triangular to stripe at $a_s/\add \approx0.770$ (cf.~Fig.~\ref{f:pd}). As $a_s$ increases further, the energy of the stripe, triangular and honeycomb states converge to the uniform state energy at the critical point. In Fig.~\ref{f:propncrit}(b) we consider the unit cell size, $L$, which was introduced earlier for the stripe state. We define the unit cell length for the triangular and honeycomb states to be\footnote{The same scaling factor can be justified for the 2D soft-core system in Ref.~\cite{Prestipino2018a}.} $L=\sqrt3 a/2$. Approaching the critical point from below, we see that $L$ converges to the roton wavelength at the critical point for all three modulated states. Note that while the triangular and honeycomb states both have a triangular unit cell, in general the honeycomb solution has a larger cell size.
The axial size $\sigma$ of the states is shown in Fig.~\ref{f:propncrit}(c). We observe that this is larger for the modulated states than the uniform state. This occurs because the modulated states are more susceptible to magnetostriction, where the DDIs cause the system to expand along the direction of the dipoles.

In Fig.~\ref{f:multiplot} we consider the behavior of the solutions at densities below and above the critical density, where the transition at the melting line is first order. For the lower density case [Figs.~\ref{f:multiplot}(a), (b), (c)] the triangular state is the ground state up to the melting point. The stripe and honeycomb states are metastable, but both of these states continuously transition to the uniform state with contrast $\mathcal{C}\to0$ and superfluid fraction $f_s\to1$. This transition occurs as $a_s\to a_{s,\mathrm{rot}}$ and with unit cell size $L$ converging to the roton wavelength. 

For the higher density case [Figs.~\ref{f:multiplot}(d), (e), (f)] the honeycomb state is the ground state up to the melting point. Here, the stripe and triangular states are metastable, and both continuously transition to the uniform state as $a_s\to a_{s,\mathrm{rot}}$, and their unit cell sizes approach $\lambda_\mathrm{rot}$. 
For the range of densities considered in the phase diagram of Fig.~\ref{f:pd} we find that the behavior discussed above is general: the metastable modulated states continuously transition at the roton instability with a unit cell length of $\lambda_\mathrm{rot}$. Related behavior is also observed in the supersolid transition in a 2D soft-core system \cite{Prestipino2018a}. For this system the triangular state is always the modulated ground-state, but stripe, honeycomb and square metastable states have a continuous transition at the roton instability point, with a unit cell size of $\lambda_\mathrm{rot}$.

In Figs.~\ref{f:multiplot}(a), (b), (c) we observe that the honeycomb solution changes character at $\as\approx0.76\add$. As $\as$ increases through this value, the contrast and unit cell size rapidly drop and the superfluid fraction saturates to unity. The insets to Figs.~\ref{f:multiplot}(b) show the honeycomb density profile either side of this transition. For $\as<0.76\add$ the honeycomb state is composed of relatively isolated high-density droplets, and the repulsion between these favors a larger unit cell size. For $\as>0.76\add$ the state instead is composed of connected hexagonal rings of moderate density (with little variation around the ring), which is conducive of the system exhibiting a high superfluid fraction.

\begin{figure}[tbh!]
	\centering
    \includegraphics[width=\linewidth]{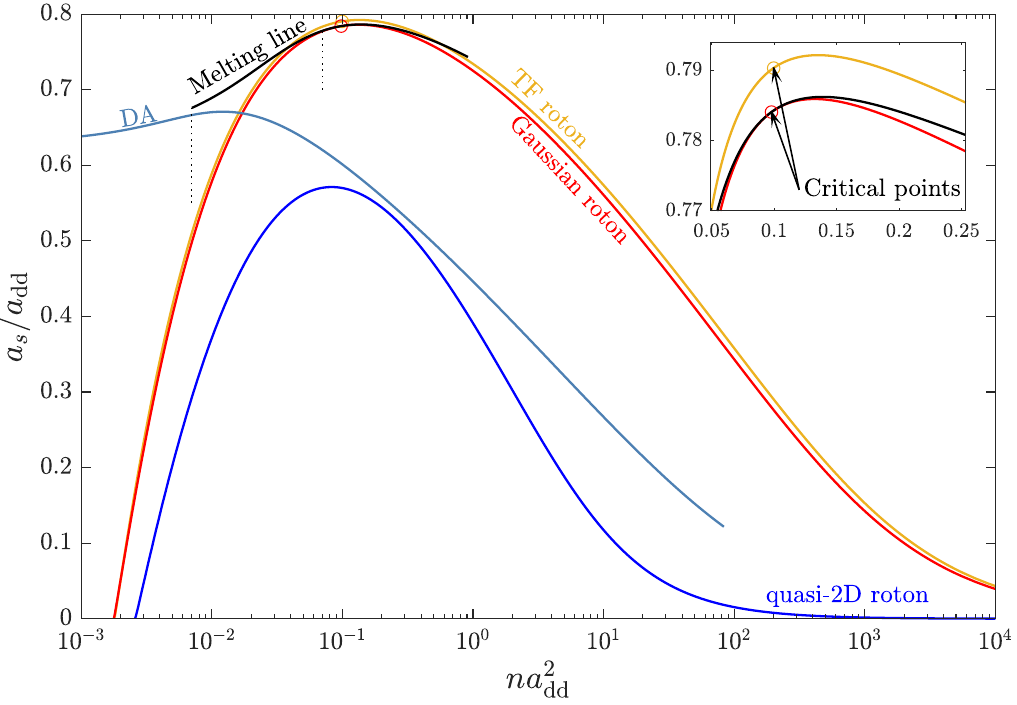}
    \caption{Roton boundaries (i.e., $\as=a_{s,\mathrm{rot}}$) across a wide range of densities for different $z$-profiles, also showing the boundary at which the energy of the uniform state exceeds the droplet ansatz (DA), with circles identifying the critical points. The inset shows a magnified view in the vicinity of the Gaussian and Thomas-Fermi (TF) critical points. Vertical dotted lines show the cases considered in Fig.~\ref{f:da}.}
	\label{f:rdawide}
\end{figure}

\section{Alternative theories and extending the phase diagram to low density}

The previous results in the paper have been calculated with the Gaussian variational ansatz for the axial wave function, and a full numerical description of the planar wave function. This has enabled us to explore the phase diagram near the critical point. In this section we first examine how changing the treatment of the axial wave function to the Thomas-Fermi approximation affects the phase diagram. We then develop two analytical treatments for describing the planar field, which provide descriptions of the stripe phase and the low density droplet array. These theories help predict the phase diagram over a broader density regime.

\subsection{Thomas-Fermi treatment of $\chi(z)$}\label{sec:TF}
For the Thomas-Fermi ansatz for the axial wave function, we set the axial field to
\begin{align}
|\chi_Z(z)|^2 = 3\max\left[\frac{1-z^2/Z^2}{4Z},0\right],
\end{align} 
with variational width $Z$.
This approach modifies the variational Gaussian treatment of the ground state (Sec.~\ref{Sec:UCtreatment}) and excitation dispersion relation (Sec.~\ref{Sec:roton}) by the following changes in parameters: $\ME_z=(Z/a_z)^2/10$, $\gamma_4=3/5Z$, $\gamma_5=45\sqrt3\pi/512Z^{3/2}$, and $G_0$ is replaced by Eq.~(36) of \cite{Baillie2015b}. 

 In Fig.~\ref{f:rdawide} we show the roton softening computed using both the Gaussian and Thomas-Fermi axial treatments. These lines are in reasonably good agreement, but give a picture of the sensitivity of our results to the axial treatment. To put these results into perspective we can also consider the quasi-2D approximation in which we assume the axial mode is the harmonic oscillator ground state (i.e.,~$\sigma=a_z$). The quasi-2D roton line lies far below the other results, emphasizing that interaction effects in the confined direction are important in a quantitative description of this system. 
  
 We also compare the effect on the critical point, determined from analysis of the modulated states [see Fig.~\ref{f:rdawide}, inset]. In the Thomas-Fermi treatment this is at $(n_\mathrm{crit}\add^2,a_{s,\mathrm{crit}}/a_{\mathrm{dd}})\approx(0.0996,0.790)$, which is in slightly better agreement with the value in Ref.~\cite{Zhang2019a} computed without a variational approximation in the axial direction [see Sec.~\ref{sec:pd}].

\subsection{CM ansatz and stripe phase properties}\label{Sec:CM}
It is feasible to produce analytical treatments of the planar field $\psi(\bm{\rho})$ for low contrast states which can be expanded in plane waves using a few of the smallest reciprocal lattice wave vectors (e.g.,~see \cite{Lu2015a}). Here we consider such an approach for the simplest case of the stripe phase by introducing the cosine-modulated (CM) ansatz: 
\begin{equation}\label{e:CM}
	\psiCM(x) = \cos\theta+\sqrt2\sin\theta\cos(2\pi x/L ),
\end{equation}
as in Refs.~\cite{Lu2015a,Blakie2020b}. Here $\theta$ is a parameter which controls the amount of modulation\footnote{Note the restriction of $\theta$ to the range $\theta\in[0,\varphi]$ for $\varphi\equiv\cot^{-1}\sqrt{2}\approx 0.616$, such that $|\psiCM|^2$ only exhibits one maximum per unit cell.}, with the state being uniform for $\theta=0$.
The planar energy per particle is
\begin{align}
    \ME_\perp &= \frac{h^2\sin^2\theta}{2mL^2} + \frac25\gQF n^{3/2}\Lambda(\theta) + \frac12\gamma_4n[g_s+2\gdd] c(\theta) \notag \\
	&+\frac32\gamma_4n\gdd \biggl[\sin^2 2\theta G_0(q_x) + \frac12\sin^4\theta G_0(2q_x)\biggr],
\end{align}
where $q_x = \sqrt2 \pi \sigma/L$, $c(\theta) = (27-4\cos2\theta-7\cos4\theta)/16$ and $\Lambda(\theta) = (90\cos\theta-55\cos3\theta-3\cos5\theta)/32$.
The variational solutions can be determined by numerically minimizing the full energy function $\mathcal{E}_\mathrm{CM}(\sigma,L ,\theta)=\mathcal{E}_z+\mathcal{E}_\perp$. The CM ansatz provides a reasonable description of the stripe state at low-to-moderate values of modulation [see Fig.~\ref{f:da}(b)]. For higher modulations (e.g.,~$\mathcal{C}\gtrsim0.5$) plane waves of higher order reciprocal vectors become important and the CM ansatz begins to differ from the numerical calculated stripe state (also see \cite{Blakie2020b,Ilg2023a}).

The CM ansatz also gives us some insight on the continuous transition into the stripe phase. Near the transition point $\theta$ is small, and to leading order in $\theta$ the stationary condition is
\begin{align}
	\frac{d\ME_\mathrm{CM}}{d\theta} = 2\theta \Delta(2\pi/L ),
	\end{align}
 where $ \Delta$ was introduced in Eq.~(\ref{e:Delta}).
The stationary points from this result are either the trivial uniform case $\theta=0$, or the modulated case with $\theta\neq0$ and $L $ determined by $\Delta(2\pi/L )=0$. The Bogoliubov spectrum for the excitations for the uniform condensate $\theta=0$ is given by Eq.~(\ref{edispersion}), and so the condition $\Delta(2\pi/L )=0$ means an excitation of wavelength $L $ has zero energy and hence we identify this as $\lambda_{\mathrm{rot}}$.

At sufficiently low $(na_{\mathrm{dd}}^2\lesssim0.012$) and high $(na_{\mathrm{dd}}^2\gtrsim0.32$) densities we find that the transition from the stripe state to the uniform state becomes discontinuous. 
However, in these regimes the CM theory predicts a continuous transition that coincides with the roton softening line. This indicates that the discontinuous transition arises from higher harmonics of the spatial wave vector $2\pi/L$ that are included in the stripe solution.

\subsection{Droplet ansatz}
Another useful approach can be developed for the modulated state based on a localized Gaussian (droplet) function in each unit cell. We refer to this as the droplet ansatz, which is generally applicable to the low density regime\footnote{This can also apply to higher densities for sufficiently low values of $\as/\add$ such that small dense (well-separated) droplets form.} (cf.~Refs.~\cite{Sepulveda2008a,Baillie2018a,Blakie2020b}). 
Here we set
\begin{align}
    \psiDA(\brho) = \sqrt{\frac{A}{\pi a_\rho^2}}e^{-\rho^2/2a_\rho^2},
\end{align}
where the variational Gaussian width $a_\rho$ should be much smaller than $a$ for consistency (cf.~Ref.~\cite{Prestipino2019a}). The planar energy per particle is then 
\begin{align}
    \ME_\perp &= \frac{\hbar^2}{2ma_\rho^2} + \frac{\gamma_4c^2}4\Bigl[g_s\!-\!\gdd f\Bigl(\frac{a_\rho}\sigma\Bigr)\Bigr] \!+\! \frac{3 sn\gdd }{8\pi \sqrt{A}} + \frac{4\gQF c^3}{25},
\end{align}
with $c=\sqrt{nA/\pi a_\rho^2}$, $f$ as defined in \cite{Lima2010a}, and 
\begin{align}
    s &= \biggl(\frac{\sqrt3}2\biggr)^{3/2}\sum_{n_1,n_2=-\infty}^\infty(n_1^2+n_1n_2+n_2^2)^{-3/2}, \\
    &= \frac{3^{1/4}}{\sqrt2} \zeta\biggl(\frac32,0\biggr)\biggl[\zeta\biggl(\frac32,\frac13\biggr)-\zeta\biggl(\frac32,\frac23\biggr)\biggr] \approx 8.89,
\end{align}
where $\zeta(3/2,a)$ is the Hurwitz zeta function.

\begin{figure}[htb!]
	\centering
    \includegraphics[width=\linewidth]{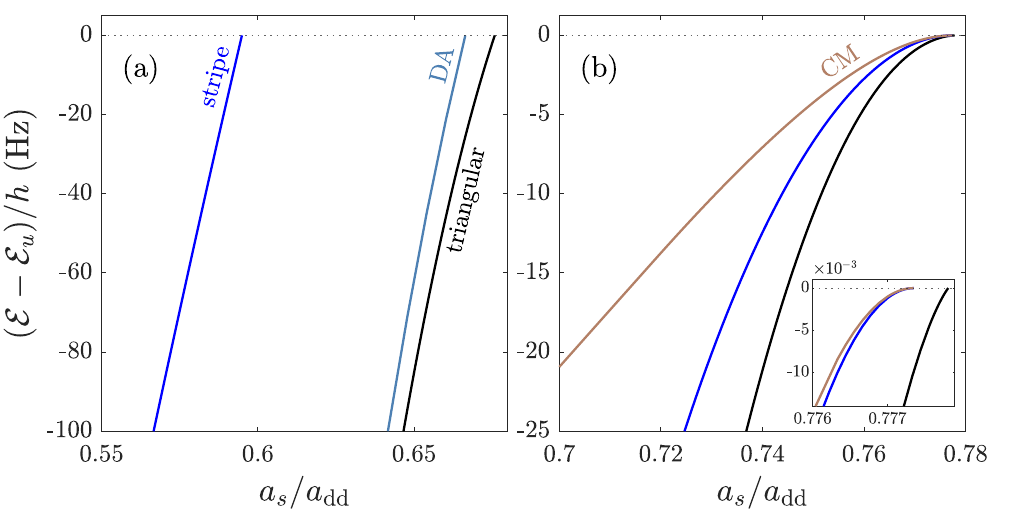} 
    \caption{Energy per particle approaching the melting line for (a) $n\add^2=\num{7.04e-3}$ comparing to DA (CM melts at $a_s/a_\mathrm{dd} = 0.498$), and (b) $n\add^2=\num{7.04e-2}$ comparing CM and stripe (DA melts at $a_s/a_\mathrm{dd} = 0.620$).} 
	\label{f:da}
\end{figure}

The results in Fig.~\ref{f:da}(a) compare the droplet ansatz to the general numerical solution for a low density case. In this regime it is energetically favorable for the atoms to form well-separated droplets, and the droplet ansatz provides a better description of the numerical solution relative to the CM ansatz. The melting point of the droplet ansatz occurs when the droplet energy exceeds the uniform background state, and the droplet is metastable. 

\subsection{Extending the melting line to the low density regime}
In Fig.~\ref{f:rdawide} we show the characteristics of the phase diagram over a much broader density range. We have applied our main numerical method to extend the melting line. However, these calculations become computationally demanding at low densities, and we turn to the CM and droplet ansatz treatments presented in this section to understand that basic behavior. 

At moderate densities in the vicinity of the critical point, the roton softening line provides a reasonable estimate of the melting line (and corresponds to the CM melting line). This can be determined by solving $\Delta=0$ from Eq.~(\ref{e:Delta}). As noted in Fig.~\ref{f:pd}, the roton line is generally slightly below the melting line, except at the critical point.

As the density decreases the roton line drops rather rapidly and for $n\add^2\lesssim2\times10^{-2}$ the droplet ansatz melting line is higher. This indicates that at low densities the droplet states are more stable than CM states at high $a_s$ values, and thus provide a better estimate for the melting line. In this regime the modulated state is formed of relatively isolated (well-separated) droplets, and the melting occurs as the droplets individually unbind. As the density decreases this unbinding levels off towards the value of a single droplet in the planar potential. Our full numerical result for the melting line extends to a minimum density of $n\add^2=7.04\times10^{-3}$, and is seen to cross over from being well described by the roton line for the moderate and high density region, to approaching the droplet ansatz melting line at low densities.

\section{Conclusions}\label{s:conlcusions}

In this work we have investigated various density modulated stationary states of the planar dipolar Bose gas in the thermodynamic limit, and obtained a ground state phase diagram. The critical point of the phase diagram occurs at an intermediate density where the uniform superfluid and the triangular, stripe, and honeycomb supersolid states all meet. Our main results use a variational Gaussian treatment that reduces the computational complexity, and we have made comparisons to a Thomas-Fermi treatment and find good agreement.
Comparison to the full 3D result in Ref.~\cite{Zhang2019a} shows both ansatzes to be qualitatively correct.
We analyze the behavior of the metastable modulated states, finding that for moderate densities these tend to transition continuously to the uniform state with a unit cell size set by the roton wavelength at the transition point. In contrast, the ground modulated state generally has a discontinuous transition to the uniform state except at the critical point.

A feature of our work is that we directly compute the superfluid fraction tensor of the 2D crystalline states. Our results show that the superfluidity is isotropic for the triangular and honeycomb phases. Interestingly, for the linear phase the superfluidity is anisotropic, with reduced superfluid fraction along the direction of modulation and full superfluidity in the perpendicular direction. 
We note that anisotropic superfluidity has been previously explored in dipolar BECs, but using a tilted dipole moment to impose anisotropic interactions \cite{Ticknor2011a,Wenzel2018a} (also see \cite{Bombin2017a,Cinti2019a,Cinti2020a,Bombin2020a}). More recently, experiments have used an optical lattice to impose a stripe pattern on non-dipolar BECs, and subsequently measure the anisotropic superfluidity \cite{Chauveau2023a,Tao2023a}. The stripe phase in the planar dipolar system is unique in that it spontaneously breaks the isotropy, and occurs as a ground state in the phase diagram. However, this state might be easily deformable, leading it to form labyrinth patterns, which could restore isotropic superfluidity.

The roton line we compute indicates the point of dynamic instability for the uniform BEC state and gives an estimate of the potential hysteresis that could occur in ramps across the transition, i.e., as $\as$ is reduced below the melting line, the uniform BEC state becomes metastable, but only becomes dynamically unstable when it crosses the roton line. Theoretical and experimental studies of such scattering length ramps in pancake shaped traps have revealed complicated dynamics, often causing the system to end up in an excited droplet configuration rather than evolving to the ground state (e.g.,~see \cite{Bisset2015a,Xi2016a,Blakie2016a,Wachtler2016b,Bisset2016a,Ferrier-Barbut2018a}).
Also, our results show that in many cases (particularly close to the critical point), the energy differences between the various ground states is quite small [e.g.~see Fig.~\ref{f:propncrit}(a)], which could lead to long equilibration times without dissipation or cooling \cite{Bland2022a}.

The cosine-modulated and droplet ansatzes provide simpler theories for estimating ground state properties over a wide parameter regime. Notably, the droplet ansatz allows us to predict the melting line at low densities, where the system will exist as a high contrast (low superfluid fraction) triangular crystal up to the melting line. The system we have examined here has been for $^{164}$Dy atoms with a fixed axial harmonic confinement strength, allowing our results to be directly compared to those of Ref.~\cite{Zhang2019a}. The effect of changing confinement has been considered in Ref.~\cite{Zhang2021a}. 

 There remain many topics for further investigation in this system, such as the dynamical stability of the ground states, which could be characterized by the excitation spectrum. Also, the possibility for phases to coexist near the first order transition boundaries.

\vspace*{0.5cm}
\noindent\textit{Note added.} In the late stages of preparation, Ref.~\cite{Zhang2023a} appeared which also shows a striped phase and discusses alternative approximations for the weakly modulated planar field. After submission, Ref.~\cite{Zhang2023b} appeared which discusses other metastable states in this system.

\section*{Acknowledgments}
We acknowledge NZ eScience Infrastructure (NeSI) High Performance Computing (HPC) facilities, and the Marsden Fund of New Zealand.
%\bibliography{physjabbrev,articles} 

\begin{thebibliography}{61}%
\makeatletter
\providecommand \@ifxundefined [1]{%
 \@ifx{#1\undefined}
}%
\providecommand \@ifnum [1]{%
 \ifnum #1\expandafter \@firstoftwo
 \else \expandafter \@secondoftwo
 \fi
}%
\providecommand \@ifx [1]{%
 \ifx #1\expandafter \@firstoftwo
 \else \expandafter \@secondoftwo
 \fi
}%
\providecommand \natexlab [1]{#1}%
\providecommand \enquote  [1]{``#1''}%
\providecommand \bibnamefont  [1]{#1}%
\providecommand \bibfnamefont [1]{#1}%
\providecommand \citenamefont [1]{#1}%
\providecommand \href@noop [0]{\@secondoftwo}%
\providecommand \href [0]{\begingroup \@sanitize@url \@href}%
\providecommand \@href[1]{\@@startlink{#1}\@@href}%
\providecommand \@@href[1]{\endgroup#1\@@endlink}%
\providecommand \@sanitize@url [0]{\catcode `\\12\catcode `\$12\catcode
  `\&12\catcode `\#12\catcode `\^12\catcode `\_12\catcode `\%12\relax}%
\providecommand \@@startlink[1]{}%
\providecommand \@@endlink[0]{}%
\providecommand \url  [0]{\begingroup\@sanitize@url \@url }%
\providecommand \@url [1]{\endgroup\@href {#1}{\urlprefix }}%
\providecommand \urlprefix  [0]{URL }%
\providecommand \Eprint [0]{\href }%
\providecommand \doibase [0]{https://doi.org/}%
\providecommand \selectlanguage [0]{\@gobble}%
\providecommand \bibinfo  [0]{\@secondoftwo}%
\providecommand \bibfield  [0]{\@secondoftwo}%
\providecommand \translation [1]{[#1]}%
\providecommand \BibitemOpen [0]{}%
\providecommand \bibitemStop [0]{}%
\providecommand \bibitemNoStop [0]{.\EOS\space}%
\providecommand \EOS [0]{\spacefactor3000\relax}%
\providecommand \BibitemShut  [1]{\csname bibitem#1\endcsname}%
\let\auto@bib@innerbib\@empty
%</preamble>
\bibitem [{\citenamefont {Tanzi}\ \emph {et~al.}(2019)\citenamefont {Tanzi},
  \citenamefont {Lucioni}, \citenamefont {Fam\`a}, \citenamefont {Catani},
  \citenamefont {Fioretti}, \citenamefont {Gabbanini}, \citenamefont {Bisset},
  \citenamefont {Santos},\ and\ \citenamefont {Modugno}}]{Tanzi2019a}%
  \BibitemOpen
  \bibfield  {author} {\bibinfo {author} {\bibfnamefont {L.}~\bibnamefont
  {Tanzi}}, \bibinfo {author} {\bibfnamefont {E.}~\bibnamefont {Lucioni}},
  \bibinfo {author} {\bibfnamefont {F.}~\bibnamefont {Fam\`a}}, \bibinfo
  {author} {\bibfnamefont {J.}~\bibnamefont {Catani}}, \bibinfo {author}
  {\bibfnamefont {A.}~\bibnamefont {Fioretti}}, \bibinfo {author}
  {\bibfnamefont {C.}~\bibnamefont {Gabbanini}}, \bibinfo {author}
  {\bibfnamefont {R.~N.}\ \bibnamefont {Bisset}}, \bibinfo {author}
  {\bibfnamefont {L.}~\bibnamefont {Santos}},\ and\ \bibinfo {author}
  {\bibfnamefont {G.}~\bibnamefont {Modugno}},\ }\bibfield  {title} {\bibinfo
  {title} {Observation of a dipolar quantum gas with metastable supersolid
  properties},\ }\href {https://doi.org/10.1103/PhysRevLett.122.130405}
  {\bibfield  {journal} {\bibinfo  {journal} {Phys. Rev. Lett.}\ }\textbf
  {\bibinfo {volume} {122}},\ \bibinfo {pages} {130405} (\bibinfo {year}
  {2019})}\BibitemShut {NoStop}%
\bibitem [{\citenamefont {B\"ottcher}\ \emph {et~al.}(2019)\citenamefont
  {B\"ottcher}, \citenamefont {Schmidt}, \citenamefont {Wenzel}, \citenamefont
  {Hertkorn}, \citenamefont {Guo}, \citenamefont {Langen},\ and\ \citenamefont
  {Pfau}}]{Bottcher2019a}%
  \BibitemOpen
  \bibfield  {author} {\bibinfo {author} {\bibfnamefont {F.}~\bibnamefont
  {B\"ottcher}}, \bibinfo {author} {\bibfnamefont {J.-N.}\ \bibnamefont
  {Schmidt}}, \bibinfo {author} {\bibfnamefont {M.}~\bibnamefont {Wenzel}},
  \bibinfo {author} {\bibfnamefont {J.}~\bibnamefont {Hertkorn}}, \bibinfo
  {author} {\bibfnamefont {M.}~\bibnamefont {Guo}}, \bibinfo {author}
  {\bibfnamefont {T.}~\bibnamefont {Langen}},\ and\ \bibinfo {author}
  {\bibfnamefont {T.}~\bibnamefont {Pfau}},\ }\bibfield  {title} {\bibinfo
  {title} {Transient supersolid properties in an array of dipolar quantum
  droplets},\ }\href {https://doi.org/10.1103/PhysRevX.9.011051} {\bibfield
  {journal} {\bibinfo  {journal} {Phys. Rev. X}\ }\textbf {\bibinfo {volume}
  {9}},\ \bibinfo {pages} {011051} (\bibinfo {year} {2019})}\BibitemShut
  {NoStop}%
\bibitem [{\citenamefont {Chomaz}\ \emph {et~al.}(2019)\citenamefont {Chomaz},
  \citenamefont {Petter}, \citenamefont {Ilzh\"ofer}, \citenamefont {Natale},
  \citenamefont {Trautmann}, \citenamefont {Politi}, \citenamefont
  {Durastante}, \citenamefont {van Bijnen}, \citenamefont {Patscheider},
  \citenamefont {Sohmen}, \citenamefont {Mark},\ and\ \citenamefont
  {Ferlaino}}]{Chomaz2019a}%
  \BibitemOpen
  \bibfield  {author} {\bibinfo {author} {\bibfnamefont {L.}~\bibnamefont
  {Chomaz}}, \bibinfo {author} {\bibfnamefont {D.}~\bibnamefont {Petter}},
  \bibinfo {author} {\bibfnamefont {P.}~\bibnamefont {Ilzh\"ofer}}, \bibinfo
  {author} {\bibfnamefont {G.}~\bibnamefont {Natale}}, \bibinfo {author}
  {\bibfnamefont {A.}~\bibnamefont {Trautmann}}, \bibinfo {author}
  {\bibfnamefont {C.}~\bibnamefont {Politi}}, \bibinfo {author} {\bibfnamefont
  {G.}~\bibnamefont {Durastante}}, \bibinfo {author} {\bibfnamefont {R.~M.~W.}\
  \bibnamefont {van Bijnen}}, \bibinfo {author} {\bibfnamefont
  {A.}~\bibnamefont {Patscheider}}, \bibinfo {author} {\bibfnamefont
  {M.}~\bibnamefont {Sohmen}}, \bibinfo {author} {\bibfnamefont {M.~J.}\
  \bibnamefont {Mark}},\ and\ \bibinfo {author} {\bibfnamefont
  {F.}~\bibnamefont {Ferlaino}},\ }\bibfield  {title} {\bibinfo {title}
  {Long-lived and transient supersolid behaviors in dipolar quantum gases},\
  }\href {https://doi.org/10.1103/PhysRevX.9.021012} {\bibfield  {journal}
  {\bibinfo  {journal} {Phys. Rev. X}\ }\textbf {\bibinfo {volume} {9}},\
  \bibinfo {pages} {021012} (\bibinfo {year} {2019})}\BibitemShut {NoStop}%
\bibitem [{\citenamefont {Norcia}\ \emph {et~al.}(2021)\citenamefont {Norcia},
  \citenamefont {Politi}, \citenamefont {Klaus}, \citenamefont {Poli},
  \citenamefont {Sohmen}, \citenamefont {Mark}, \citenamefont {Bisset},
  \citenamefont {Santos},\ and\ \citenamefont {Ferlaino}}]{Norcia2021a}%
  \BibitemOpen
  \bibfield  {author} {\bibinfo {author} {\bibfnamefont {M.~A.}\ \bibnamefont
  {Norcia}}, \bibinfo {author} {\bibfnamefont {C.}~\bibnamefont {Politi}},
  \bibinfo {author} {\bibfnamefont {L.}~\bibnamefont {Klaus}}, \bibinfo
  {author} {\bibfnamefont {E.}~\bibnamefont {Poli}}, \bibinfo {author}
  {\bibfnamefont {M.}~\bibnamefont {Sohmen}}, \bibinfo {author} {\bibfnamefont
  {M.~J.}\ \bibnamefont {Mark}}, \bibinfo {author} {\bibfnamefont {R.~N.}\
  \bibnamefont {Bisset}}, \bibinfo {author} {\bibfnamefont {L.}~\bibnamefont
  {Santos}},\ and\ \bibinfo {author} {\bibfnamefont {F.}~\bibnamefont
  {Ferlaino}},\ }\bibfield  {title} {\bibinfo {title} {Two-dimensional
  supersolidity in a dipolar quantum gas},\ }\href
  {https://doi.org/10.1038/s41586-021-03725-7} {\bibfield  {journal} {\bibinfo
  {journal} {Nature}\ }\textbf {\bibinfo {volume} {596}},\ \bibinfo {pages}
  {357} (\bibinfo {year} {2021})}\BibitemShut {NoStop}%
\bibitem [{\citenamefont {Chomaz}\ \emph {et~al.}(2022)\citenamefont {Chomaz},
  \citenamefont {Ferrier-Barbut}, \citenamefont {Ferlaino}, \citenamefont
  {Laburthe-Tolra}, \citenamefont {Lev},\ and\ \citenamefont
  {Pfau}}]{Chomaz2023a}%
  \BibitemOpen
  \bibfield  {author} {\bibinfo {author} {\bibfnamefont {L.}~\bibnamefont
  {Chomaz}}, \bibinfo {author} {\bibfnamefont {I.}~\bibnamefont
  {Ferrier-Barbut}}, \bibinfo {author} {\bibfnamefont {F.}~\bibnamefont
  {Ferlaino}}, \bibinfo {author} {\bibfnamefont {B.}~\bibnamefont
  {Laburthe-Tolra}}, \bibinfo {author} {\bibfnamefont {B.~L.}\ \bibnamefont
  {Lev}},\ and\ \bibinfo {author} {\bibfnamefont {T.}~\bibnamefont {Pfau}},\
  }\bibfield  {title} {\bibinfo {title} {Dipolar physics: a review of
  experiments with magnetic quantum gases},\ }\href
  {https://doi.org/10.1088/1361-6633/aca814} {\bibfield  {journal} {\bibinfo
  {journal} {Rep. Prog. Phys.}\ }\textbf {\bibinfo {volume} {86}},\ \bibinfo
  {pages} {026401} (\bibinfo {year} {2022})}\BibitemShut {NoStop}%
\bibitem [{\citenamefont {Roccuzzo}\ and\ \citenamefont
  {Ancilotto}(2019)}]{Roccuzzo2019a}%
  \BibitemOpen
  \bibfield  {author} {\bibinfo {author} {\bibfnamefont {S.~M.}\ \bibnamefont
  {Roccuzzo}}\ and\ \bibinfo {author} {\bibfnamefont {F.}~\bibnamefont
  {Ancilotto}},\ }\bibfield  {title} {\bibinfo {title} {Supersolid behavior of
  a dipolar {Bose}-{Einstein} condensate confined in a tube},\ }\href
  {https://doi.org/10.1103/PhysRevA.99.041601} {\bibfield  {journal} {\bibinfo
  {journal} {Phys. Rev. A}\ }\textbf {\bibinfo {volume} {99}},\ \bibinfo
  {pages} {041601} (\bibinfo {year} {2019})}\BibitemShut {NoStop}%
\bibitem [{\citenamefont {Blakie}\ \emph {et~al.}(2020)\citenamefont {Blakie},
  \citenamefont {Baillie}, \citenamefont {Chomaz},\ and\ \citenamefont
  {Ferlaino}}]{Blakie2020b}%
  \BibitemOpen
  \bibfield  {author} {\bibinfo {author} {\bibfnamefont {P.~B.}\ \bibnamefont
  {Blakie}}, \bibinfo {author} {\bibfnamefont {D.}~\bibnamefont {Baillie}},
  \bibinfo {author} {\bibfnamefont {L.}~\bibnamefont {Chomaz}},\ and\ \bibinfo
  {author} {\bibfnamefont {F.}~\bibnamefont {Ferlaino}},\ }\bibfield  {title}
  {\bibinfo {title} {Supersolidity in an elongated dipolar condensate},\ }\href
  {https://doi.org/10.1103/PhysRevResearch.2.043318} {\bibfield  {journal}
  {\bibinfo  {journal} {Phys. Rev. Research}\ }\textbf {\bibinfo {volume}
  {2}},\ \bibinfo {pages} {043318} (\bibinfo {year} {2020})}\BibitemShut
  {NoStop}%
\bibitem [{\citenamefont {Biagioni}\ \emph {et~al.}(2022)\citenamefont
  {Biagioni}, \citenamefont {Antolini}, \citenamefont {Ala\~na}, \citenamefont
  {Modugno}, \citenamefont {Fioretti}, \citenamefont {Gabbanini}, \citenamefont
  {Tanzi},\ and\ \citenamefont {Modugno}}]{Biagioni2022a}%
  \BibitemOpen
  \bibfield  {author} {\bibinfo {author} {\bibfnamefont {G.}~\bibnamefont
  {Biagioni}}, \bibinfo {author} {\bibfnamefont {N.}~\bibnamefont {Antolini}},
  \bibinfo {author} {\bibfnamefont {A.}~\bibnamefont {Ala\~na}}, \bibinfo
  {author} {\bibfnamefont {M.}~\bibnamefont {Modugno}}, \bibinfo {author}
  {\bibfnamefont {A.}~\bibnamefont {Fioretti}}, \bibinfo {author}
  {\bibfnamefont {C.}~\bibnamefont {Gabbanini}}, \bibinfo {author}
  {\bibfnamefont {L.}~\bibnamefont {Tanzi}},\ and\ \bibinfo {author}
  {\bibfnamefont {G.}~\bibnamefont {Modugno}},\ }\bibfield  {title} {\bibinfo
  {title} {Dimensional crossover in the superfluid-supersolid quantum phase
  transition},\ }\href {https://doi.org/10.1103/PhysRevX.12.021019} {\bibfield
  {journal} {\bibinfo  {journal} {Phys. Rev. X}\ }\textbf {\bibinfo {volume}
  {12}},\ \bibinfo {pages} {021019} (\bibinfo {year} {2022})}\BibitemShut
  {NoStop}%
\bibitem [{\citenamefont {Ilg}\ and\ \citenamefont
  {B\"uchler}(2023)}]{Ilg2023a}%
  \BibitemOpen
  \bibfield  {author} {\bibinfo {author} {\bibfnamefont {T.}~\bibnamefont
  {Ilg}}\ and\ \bibinfo {author} {\bibfnamefont {H.~P.}\ \bibnamefont
  {B\"uchler}},\ }\bibfield  {title} {\bibinfo {title} {Ground-state stability
  and excitation spectrum of a one-dimensional dipolar supersolid},\ }\href
  {https://doi.org/10.1103/PhysRevA.107.013314} {\bibfield  {journal} {\bibinfo
   {journal} {Phys. Rev. A}\ }\textbf {\bibinfo {volume} {107}},\ \bibinfo
  {pages} {013314} (\bibinfo {year} {2023})}\BibitemShut {NoStop}%
\bibitem [{\citenamefont {Smith}\ \emph {et~al.}(2023)\citenamefont {Smith},
  \citenamefont {Baillie},\ and\ \citenamefont {Blakie}}]{Smith2023a}%
  \BibitemOpen
  \bibfield  {author} {\bibinfo {author} {\bibfnamefont {J.~C.}\ \bibnamefont
  {Smith}}, \bibinfo {author} {\bibfnamefont {D.}~\bibnamefont {Baillie}},\
  and\ \bibinfo {author} {\bibfnamefont {P.~B.}\ \bibnamefont {Blakie}},\
  }\bibfield  {title} {\bibinfo {title} {Supersolidity and crystallization of a
  dipolar {Bose} gas in an infinite tube},\ }\href
  {https://doi.org/10.1103/PhysRevA.107.033301} {\bibfield  {journal} {\bibinfo
   {journal} {Phys. Rev. A}\ }\textbf {\bibinfo {volume} {107}},\ \bibinfo
  {pages} {033301} (\bibinfo {year} {2023})}\BibitemShut {NoStop}%
\bibitem [{\citenamefont {Sep\'ulveda}\ \emph {et~al.}(2008)\citenamefont
  {Sep\'ulveda}, \citenamefont {Josserand},\ and\ \citenamefont
  {Rica}}]{Sepulveda2008a}%
  \BibitemOpen
  \bibfield  {author} {\bibinfo {author} {\bibfnamefont {N.}~\bibnamefont
  {Sep\'ulveda}}, \bibinfo {author} {\bibfnamefont {C.}~\bibnamefont
  {Josserand}},\ and\ \bibinfo {author} {\bibfnamefont {S.}~\bibnamefont
  {Rica}},\ }\bibfield  {title} {\bibinfo {title} {Nonclassical rotational
  inertia fraction in a one-dimensional model of a supersolid},\ }\href
  {https://doi.org/10.1103/PhysRevB.77.054513} {\bibfield  {journal} {\bibinfo
  {journal} {Phys. Rev. B}\ }\textbf {\bibinfo {volume} {77}},\ \bibinfo
  {pages} {054513} (\bibinfo {year} {2008})}\BibitemShut {NoStop}%
\bibitem [{\citenamefont {Lu}\ \emph {et~al.}(2015)\citenamefont {Lu},
  \citenamefont {Li}, \citenamefont {Petrov},\ and\ \citenamefont
  {Shlyapnikov}}]{Lu2015a}%
  \BibitemOpen
  \bibfield  {author} {\bibinfo {author} {\bibfnamefont {Z.-K.}\ \bibnamefont
  {Lu}}, \bibinfo {author} {\bibfnamefont {Y.}~\bibnamefont {Li}}, \bibinfo
  {author} {\bibfnamefont {D.~S.}\ \bibnamefont {Petrov}},\ and\ \bibinfo
  {author} {\bibfnamefont {G.~V.}\ \bibnamefont {Shlyapnikov}},\ }\bibfield
  {title} {\bibinfo {title} {Stable dilute supersolid of two-dimensional
  dipolar bosons},\ }\href {https://doi.org/10.1103/PhysRevLett.115.075303}
  {\bibfield  {journal} {\bibinfo  {journal} {Phys. Rev. Lett.}\ }\textbf
  {\bibinfo {volume} {115}},\ \bibinfo {pages} {075303} (\bibinfo {year}
  {2015})}\BibitemShut {NoStop}%
\bibitem [{\citenamefont {Zhang}\ \emph {et~al.}(2019)\citenamefont {Zhang},
  \citenamefont {Maucher},\ and\ \citenamefont {Pohl}}]{Zhang2019a}%
  \BibitemOpen
  \bibfield  {author} {\bibinfo {author} {\bibfnamefont {Y.-C.}\ \bibnamefont
  {Zhang}}, \bibinfo {author} {\bibfnamefont {F.}~\bibnamefont {Maucher}},\
  and\ \bibinfo {author} {\bibfnamefont {T.}~\bibnamefont {Pohl}},\ }\bibfield
  {title} {\bibinfo {title} {Supersolidity around a critical point in dipolar
  {Bose}-{Einstein} condensates},\ }\href
  {https://doi.org/10.1103/PhysRevLett.123.015301} {\bibfield  {journal}
  {\bibinfo  {journal} {Phys. Rev. Lett.}\ }\textbf {\bibinfo {volume} {123}},\
  \bibinfo {pages} {015301} (\bibinfo {year} {2019})}\BibitemShut {NoStop}%
\bibitem [{\citenamefont {Watanabe}\ and\ \citenamefont
  {Murayama}(2012)}]{Watanabe2012a}%
  \BibitemOpen
  \bibfield  {author} {\bibinfo {author} {\bibfnamefont {H.}~\bibnamefont
  {Watanabe}}\ and\ \bibinfo {author} {\bibfnamefont {H.}~\bibnamefont
  {Murayama}},\ }\bibfield  {title} {\bibinfo {title} {Unified description of
  {Nambu}-{Goldstone} bosons without {Lorentz} invariance},\ }\href
  {https://doi.org/10.1103/PhysRevLett.108.251602} {\bibfield  {journal}
  {\bibinfo  {journal} {Phys. Rev. Lett.}\ }\textbf {\bibinfo {volume} {108}},\
  \bibinfo {pages} {251602} (\bibinfo {year} {2012})}\BibitemShut {NoStop}%
\bibitem [{\citenamefont {Klaus}\ \emph {et~al.}(2022)\citenamefont {Klaus},
  \citenamefont {Bland}, \citenamefont {Poli}, \citenamefont {Politi},
  \citenamefont {Lamporesi}, \citenamefont {Casotti}, \citenamefont {Bisset},
  \citenamefont {Mark},\ and\ \citenamefont {Ferlaino}}]{Klaus2022a}%
  \BibitemOpen
  \bibfield  {author} {\bibinfo {author} {\bibfnamefont {L.}~\bibnamefont
  {Klaus}}, \bibinfo {author} {\bibfnamefont {T.}~\bibnamefont {Bland}},
  \bibinfo {author} {\bibfnamefont {E.}~\bibnamefont {Poli}}, \bibinfo {author}
  {\bibfnamefont {C.}~\bibnamefont {Politi}}, \bibinfo {author} {\bibfnamefont
  {G.}~\bibnamefont {Lamporesi}}, \bibinfo {author} {\bibfnamefont
  {E.}~\bibnamefont {Casotti}}, \bibinfo {author} {\bibfnamefont {R.~N.}\
  \bibnamefont {Bisset}}, \bibinfo {author} {\bibfnamefont {M.~J.}\
  \bibnamefont {Mark}},\ and\ \bibinfo {author} {\bibfnamefont
  {F.}~\bibnamefont {Ferlaino}},\ }\bibfield  {title} {\bibinfo {title}
  {Observation of vortices and vortex stripes in a dipolar condensate},\ }\href
  {https://doi.org/10.1038/s41567-022-01793-8} {\bibfield  {journal} {\bibinfo
  {journal} {Nat. Phys.}\ }\textbf {\bibinfo {volume} {18}},\ \bibinfo {pages}
  {1453} (\bibinfo {year} {2022})}\BibitemShut {NoStop}%
\bibitem [{\citenamefont {Pomeau}\ and\ \citenamefont
  {Rica}(1994)}]{Pomeau1994a}%
  \BibitemOpen
  \bibfield  {author} {\bibinfo {author} {\bibfnamefont {Y.}~\bibnamefont
  {Pomeau}}\ and\ \bibinfo {author} {\bibfnamefont {S.}~\bibnamefont {Rica}},\
  }\bibfield  {title} {\bibinfo {title} {Dynamics of a model of supersolid},\
  }\href {https://doi.org/10.1103/PhysRevLett.72.2426} {\bibfield  {journal}
  {\bibinfo  {journal} {Phys. Rev. Lett.}\ }\textbf {\bibinfo {volume} {72}},\
  \bibinfo {pages} {2426} (\bibinfo {year} {1994})}\BibitemShut {NoStop}%
\bibitem [{\citenamefont {Saccani}\ \emph {et~al.}(2012)\citenamefont
  {Saccani}, \citenamefont {Moroni},\ and\ \citenamefont
  {Boninsegni}}]{Saccani2012a}%
  \BibitemOpen
  \bibfield  {author} {\bibinfo {author} {\bibfnamefont {S.}~\bibnamefont
  {Saccani}}, \bibinfo {author} {\bibfnamefont {S.}~\bibnamefont {Moroni}},\
  and\ \bibinfo {author} {\bibfnamefont {M.}~\bibnamefont {Boninsegni}},\
  }\bibfield  {title} {\bibinfo {title} {Excitation spectrum of a supersolid},\
  }\href {https://doi.org/10.1103/PhysRevLett.108.175301} {\bibfield  {journal}
  {\bibinfo  {journal} {Phys. Rev. Lett.}\ }\textbf {\bibinfo {volume} {108}},\
  \bibinfo {pages} {175301} (\bibinfo {year} {2012})}\BibitemShut {NoStop}%
\bibitem [{\citenamefont {Hsueh}\ \emph {et~al.}(2012)\citenamefont {Hsueh},
  \citenamefont {Lin}, \citenamefont {Horng},\ and\ \citenamefont
  {Wu}}]{Hsueh2012a}%
  \BibitemOpen
  \bibfield  {author} {\bibinfo {author} {\bibfnamefont {C.-H.}\ \bibnamefont
  {Hsueh}}, \bibinfo {author} {\bibfnamefont {T.-C.}\ \bibnamefont {Lin}},
  \bibinfo {author} {\bibfnamefont {T.-L.}\ \bibnamefont {Horng}},\ and\
  \bibinfo {author} {\bibfnamefont {W.~C.}\ \bibnamefont {Wu}},\ }\bibfield
  {title} {\bibinfo {title} {Quantum crystals in a trapped {Rydberg}-dressed
  {Bose}-{Einstein} condensate},\ }\href
  {https://doi.org/10.1103/PhysRevA.86.013619} {\bibfield  {journal} {\bibinfo
  {journal} {Phys. Rev. A}\ }\textbf {\bibinfo {volume} {86}},\ \bibinfo
  {pages} {013619} (\bibinfo {year} {2012})}\BibitemShut {NoStop}%
\bibitem [{\citenamefont {Kunimi}\ and\ \citenamefont
  {Kato}(2012)}]{Kunimi2012a}%
  \BibitemOpen
  \bibfield  {author} {\bibinfo {author} {\bibfnamefont {M.}~\bibnamefont
  {Kunimi}}\ and\ \bibinfo {author} {\bibfnamefont {Y.}~\bibnamefont {Kato}},\
  }\bibfield  {title} {\bibinfo {title} {Mean-field and stability analyses of
  two-dimensional flowing soft-core bosons modeling a supersolid},\ }\href
  {https://doi.org/10.1103/PhysRevB.86.060510} {\bibfield  {journal} {\bibinfo
  {journal} {Phys. Rev. B}\ }\textbf {\bibinfo {volume} {86}},\ \bibinfo
  {pages} {060510} (\bibinfo {year} {2012})}\BibitemShut {NoStop}%
\bibitem [{\citenamefont {Macr\`{\i}}\ \emph {et~al.}(2013)\citenamefont
  {Macr\`{\i}}, \citenamefont {Maucher}, \citenamefont {Cinti},\ and\
  \citenamefont {Pohl}}]{Macri2013a}%
  \BibitemOpen
  \bibfield  {author} {\bibinfo {author} {\bibfnamefont {T.}~\bibnamefont
  {Macr\`{\i}}}, \bibinfo {author} {\bibfnamefont {F.}~\bibnamefont {Maucher}},
  \bibinfo {author} {\bibfnamefont {F.}~\bibnamefont {Cinti}},\ and\ \bibinfo
  {author} {\bibfnamefont {T.}~\bibnamefont {Pohl}},\ }\bibfield  {title}
  {\bibinfo {title} {Elementary excitations of ultracold soft-core bosons
  across the superfluid-supersolid phase transition},\ }\href
  {https://doi.org/10.1103/PhysRevA.87.061602} {\bibfield  {journal} {\bibinfo
  {journal} {Phys. Rev. A}\ }\textbf {\bibinfo {volume} {87}},\ \bibinfo
  {pages} {061602} (\bibinfo {year} {2013})}\BibitemShut {NoStop}%
\bibitem [{\citenamefont {Prestipino}\ \emph
  {et~al.}(2018{\natexlab{a}})\citenamefont {Prestipino}, \citenamefont
  {Sergi},\ and\ \citenamefont {Bruno}}]{Prestipino2018a}%
  \BibitemOpen
  \bibfield  {author} {\bibinfo {author} {\bibfnamefont {S.}~\bibnamefont
  {Prestipino}}, \bibinfo {author} {\bibfnamefont {A.}~\bibnamefont {Sergi}},\
  and\ \bibinfo {author} {\bibfnamefont {E.}~\bibnamefont {Bruno}},\ }\bibfield
   {title} {\bibinfo {title} {Freezing of soft-core bosons at zero temperature:
  A variational theory},\ }\href {https://doi.org/10.1103/PhysRevB.98.104104}
  {\bibfield  {journal} {\bibinfo  {journal} {Phys. Rev. B}\ }\textbf {\bibinfo
  {volume} {98}},\ \bibinfo {pages} {104104} (\bibinfo {year}
  {2018}{\natexlab{a}})}\BibitemShut {NoStop}%
\bibitem [{\citenamefont {Baillie}\ and\ \citenamefont
  {Blakie}(2018)}]{Baillie2018a}%
  \BibitemOpen
  \bibfield  {author} {\bibinfo {author} {\bibfnamefont {D.}~\bibnamefont
  {Baillie}}\ and\ \bibinfo {author} {\bibfnamefont {P.~B.}\ \bibnamefont
  {Blakie}},\ }\bibfield  {title} {\bibinfo {title} {Droplet crystal ground
  states of a dipolar {Bose} gas},\ }\href
  {https://doi.org/10.1103/PhysRevLett.121.195301} {\bibfield  {journal}
  {\bibinfo  {journal} {Phys. Rev. Lett.}\ }\textbf {\bibinfo {volume} {121}},\
  \bibinfo {pages} {195301} (\bibinfo {year} {2018})}\BibitemShut {NoStop}%
\bibitem [{\citenamefont {Zhang}\ \emph {et~al.}(2021)\citenamefont {Zhang},
  \citenamefont {Pohl},\ and\ \citenamefont {Maucher}}]{Zhang2021a}%
  \BibitemOpen
  \bibfield  {author} {\bibinfo {author} {\bibfnamefont {Y.-C.}\ \bibnamefont
  {Zhang}}, \bibinfo {author} {\bibfnamefont {T.}~\bibnamefont {Pohl}},\ and\
  \bibinfo {author} {\bibfnamefont {F.}~\bibnamefont {Maucher}},\ }\bibfield
  {title} {\bibinfo {title} {Phases of supersolids in confined dipolar
  {B}ose-{E}instein condensates},\ }\href
  {https://doi.org/10.1103/PhysRevA.104.013310} {\bibfield  {journal} {\bibinfo
   {journal} {Phys. Rev. A}\ }\textbf {\bibinfo {volume} {104}},\ \bibinfo
  {pages} {013310} (\bibinfo {year} {2021})}\BibitemShut {NoStop}%
\bibitem [{\citenamefont {Hertkorn}\ \emph {et~al.}(2021)\citenamefont
  {Hertkorn}, \citenamefont {Schmidt}, \citenamefont {Guo}, \citenamefont
  {B\"ottcher}, \citenamefont {Ng}, \citenamefont {Graham}, \citenamefont
  {Uerlings}, \citenamefont {Langen}, \citenamefont {Zwierlein},\ and\
  \citenamefont {Pfau}}]{Hertkorn2021a}%
  \BibitemOpen
  \bibfield  {author} {\bibinfo {author} {\bibfnamefont {J.}~\bibnamefont
  {Hertkorn}}, \bibinfo {author} {\bibfnamefont {J.-N.}\ \bibnamefont
  {Schmidt}}, \bibinfo {author} {\bibfnamefont {M.}~\bibnamefont {Guo}},
  \bibinfo {author} {\bibfnamefont {F.}~\bibnamefont {B\"ottcher}}, \bibinfo
  {author} {\bibfnamefont {K.~S.~H.}\ \bibnamefont {Ng}}, \bibinfo {author}
  {\bibfnamefont {S.~D.}\ \bibnamefont {Graham}}, \bibinfo {author}
  {\bibfnamefont {P.}~\bibnamefont {Uerlings}}, \bibinfo {author}
  {\bibfnamefont {T.}~\bibnamefont {Langen}}, \bibinfo {author} {\bibfnamefont
  {M.}~\bibnamefont {Zwierlein}},\ and\ \bibinfo {author} {\bibfnamefont
  {T.}~\bibnamefont {Pfau}},\ }\bibfield  {title} {\bibinfo {title} {Pattern
  formation in quantum ferrofluids: From supersolids to superglasses},\ }\href
  {https://doi.org/10.1103/PhysRevResearch.3.033125} {\bibfield  {journal}
  {\bibinfo  {journal} {Phys. Rev. Research}\ }\textbf {\bibinfo {volume}
  {3}},\ \bibinfo {pages} {033125} (\bibinfo {year} {2021})}\BibitemShut
  {NoStop}%
\bibitem [{\citenamefont {Poli}\ \emph {et~al.}(2021)\citenamefont {Poli},
  \citenamefont {Bland}, \citenamefont {Politi}, \citenamefont {Klaus},
  \citenamefont {Norcia}, \citenamefont {Ferlaino}, \citenamefont {Bisset},\
  and\ \citenamefont {Santos}}]{Poli2021a}%
  \BibitemOpen
  \bibfield  {author} {\bibinfo {author} {\bibfnamefont {E.}~\bibnamefont
  {Poli}}, \bibinfo {author} {\bibfnamefont {T.}~\bibnamefont {Bland}},
  \bibinfo {author} {\bibfnamefont {C.}~\bibnamefont {Politi}}, \bibinfo
  {author} {\bibfnamefont {L.}~\bibnamefont {Klaus}}, \bibinfo {author}
  {\bibfnamefont {M.~A.}\ \bibnamefont {Norcia}}, \bibinfo {author}
  {\bibfnamefont {F.}~\bibnamefont {Ferlaino}}, \bibinfo {author}
  {\bibfnamefont {R.~N.}\ \bibnamefont {Bisset}},\ and\ \bibinfo {author}
  {\bibfnamefont {L.}~\bibnamefont {Santos}},\ }\bibfield  {title} {\bibinfo
  {title} {Maintaining supersolidity in one and two dimensions},\ }\href
  {https://doi.org/10.1103/PhysRevA.104.063307} {\bibfield  {journal} {\bibinfo
   {journal} {Phys. Rev. A}\ }\textbf {\bibinfo {volume} {104}},\ \bibinfo
  {pages} {063307} (\bibinfo {year} {2021})}\BibitemShut {NoStop}%
\bibitem [{\citenamefont {Schmidt}\ \emph {et~al.}(2022)\citenamefont
  {Schmidt}, \citenamefont {Lassabli\`ere}, \citenamefont {Qu\'em\'ener},\ and\
  \citenamefont {Langen}}]{Schmidt2022a}%
  \BibitemOpen
  \bibfield  {author} {\bibinfo {author} {\bibfnamefont {M.}~\bibnamefont
  {Schmidt}}, \bibinfo {author} {\bibfnamefont {L.}~\bibnamefont
  {Lassabli\`ere}}, \bibinfo {author} {\bibfnamefont {G.}~\bibnamefont
  {Qu\'em\'ener}},\ and\ \bibinfo {author} {\bibfnamefont {T.}~\bibnamefont
  {Langen}},\ }\bibfield  {title} {\bibinfo {title} {Self-bound dipolar
  droplets and supersolids in molecular {Bose}-{Einstein} condensates},\ }\href
  {https://doi.org/10.1103/PhysRevResearch.4.013235} {\bibfield  {journal}
  {\bibinfo  {journal} {Phys. Rev. Research}\ }\textbf {\bibinfo {volume}
  {4}},\ \bibinfo {pages} {013235} (\bibinfo {year} {2022})}\BibitemShut
  {NoStop}%
\bibitem [{\citenamefont {Kadau}\ \emph {et~al.}(2016)\citenamefont {Kadau},
  \citenamefont {Schmitt}, \citenamefont {Wenzel}, \citenamefont {Wink},
  \citenamefont {Maier}, \citenamefont {Ferrier-Barbut},\ and\ \citenamefont
  {Pfau}}]{Kadau2016a}%
  \BibitemOpen
  \bibfield  {author} {\bibinfo {author} {\bibfnamefont {H.}~\bibnamefont
  {Kadau}}, \bibinfo {author} {\bibfnamefont {M.}~\bibnamefont {Schmitt}},
  \bibinfo {author} {\bibfnamefont {M.}~\bibnamefont {Wenzel}}, \bibinfo
  {author} {\bibfnamefont {C.}~\bibnamefont {Wink}}, \bibinfo {author}
  {\bibfnamefont {T.}~\bibnamefont {Maier}}, \bibinfo {author} {\bibfnamefont
  {I.}~\bibnamefont {Ferrier-Barbut}},\ and\ \bibinfo {author} {\bibfnamefont
  {T.}~\bibnamefont {Pfau}},\ }\bibfield  {title} {\bibinfo {title} {Observing
  the {Rosensweig} instability of a quantum ferrofluid},\ }\href
  {http://dx.doi.org/10.1038/nature16485} {\bibfield  {journal} {\bibinfo
  {journal} {Nature}\ }\textbf {\bibinfo {volume} {530}},\ \bibinfo {pages}
  {194} (\bibinfo {year} {2016})}\BibitemShut {NoStop}%
\bibitem [{\citenamefont {Bisset}\ and\ \citenamefont
  {Blakie}(2015)}]{Bisset2015a}%
  \BibitemOpen
  \bibfield  {author} {\bibinfo {author} {\bibfnamefont {R.~N.}\ \bibnamefont
  {Bisset}}\ and\ \bibinfo {author} {\bibfnamefont {P.~B.}\ \bibnamefont
  {Blakie}},\ }\bibfield  {title} {\bibinfo {title} {Crystallization of a
  dilute atomic dipolar condensate},\ }\href
  {https://doi.org/10.1103/PhysRevA.92.061603} {\bibfield  {journal} {\bibinfo
  {journal} {Phys. Rev. A}\ }\textbf {\bibinfo {volume} {92}},\ \bibinfo
  {pages} {061603} (\bibinfo {year} {2015})}\BibitemShut {NoStop}%
\bibitem [{\citenamefont {Xi}\ and\ \citenamefont {Saito}(2016)}]{Xi2016a}%
  \BibitemOpen
  \bibfield  {author} {\bibinfo {author} {\bibfnamefont {K.-T.}\ \bibnamefont
  {Xi}}\ and\ \bibinfo {author} {\bibfnamefont {H.}~\bibnamefont {Saito}},\
  }\bibfield  {title} {\bibinfo {title} {Droplet formation in a
  {Bose}-{Einstein} condensate with strong dipole-dipole interaction},\ }\href
  {https://doi.org/10.1103/PhysRevA.93.011604} {\bibfield  {journal} {\bibinfo
  {journal} {Phys. Rev. A}\ }\textbf {\bibinfo {volume} {93}},\ \bibinfo
  {pages} {011604} (\bibinfo {year} {2016})}\BibitemShut {NoStop}%
\bibitem [{\citenamefont {Blakie}(2016)}]{Blakie2016a}%
  \BibitemOpen
  \bibfield  {author} {\bibinfo {author} {\bibfnamefont {P.~B.}\ \bibnamefont
  {Blakie}},\ }\bibfield  {title} {\bibinfo {title} {Properties of a dipolar
  condensate with three-body interactions},\ }\href
  {https://doi.org/10.1103/PhysRevA.93.033644} {\bibfield  {journal} {\bibinfo
  {journal} {Phys. Rev. A}\ }\textbf {\bibinfo {volume} {93}},\ \bibinfo
  {pages} {033644} (\bibinfo {year} {2016})}\BibitemShut {NoStop}%
\bibitem [{\citenamefont {Ferrier-Barbut}\ \emph {et~al.}(2018)\citenamefont
  {Ferrier-Barbut}, \citenamefont {Wenzel}, \citenamefont {Schmitt},
  \citenamefont {B\"ottcher},\ and\ \citenamefont
  {Pfau}}]{Ferrier-Barbut2018a}%
  \BibitemOpen
  \bibfield  {author} {\bibinfo {author} {\bibfnamefont {I.}~\bibnamefont
  {Ferrier-Barbut}}, \bibinfo {author} {\bibfnamefont {M.}~\bibnamefont
  {Wenzel}}, \bibinfo {author} {\bibfnamefont {M.}~\bibnamefont {Schmitt}},
  \bibinfo {author} {\bibfnamefont {F.}~\bibnamefont {B\"ottcher}},\ and\
  \bibinfo {author} {\bibfnamefont {T.}~\bibnamefont {Pfau}},\ }\bibfield
  {title} {\bibinfo {title} {Onset of a modulational instability in trapped
  dipolar {Bose}-{Einstein} condensates},\ }\href
  {https://doi.org/10.1103/PhysRevA.97.011604} {\bibfield  {journal} {\bibinfo
  {journal} {Phys. Rev. A}\ }\textbf {\bibinfo {volume} {97}},\ \bibinfo
  {pages} {011604(R)} (\bibinfo {year} {2018})}\BibitemShut {NoStop}%
\bibitem [{\citenamefont {Bland}\ \emph {et~al.}(2022)\citenamefont {Bland},
  \citenamefont {Poli}, \citenamefont {Politi}, \citenamefont {Klaus},
  \citenamefont {Norcia}, \citenamefont {Ferlaino}, \citenamefont {Santos},\
  and\ \citenamefont {Bisset}}]{Bland2022a}%
  \BibitemOpen
  \bibfield  {author} {\bibinfo {author} {\bibfnamefont {T.}~\bibnamefont
  {Bland}}, \bibinfo {author} {\bibfnamefont {E.}~\bibnamefont {Poli}},
  \bibinfo {author} {\bibfnamefont {C.}~\bibnamefont {Politi}}, \bibinfo
  {author} {\bibfnamefont {L.}~\bibnamefont {Klaus}}, \bibinfo {author}
  {\bibfnamefont {M.~A.}\ \bibnamefont {Norcia}}, \bibinfo {author}
  {\bibfnamefont {F.}~\bibnamefont {Ferlaino}}, \bibinfo {author}
  {\bibfnamefont {L.}~\bibnamefont {Santos}},\ and\ \bibinfo {author}
  {\bibfnamefont {R.~N.}\ \bibnamefont {Bisset}},\ }\bibfield  {title}
  {\bibinfo {title} {Two-dimensional supersolid formation in dipolar
  condensates},\ }\href {https://doi.org/10.1103/PhysRevLett.128.195302}
  {\bibfield  {journal} {\bibinfo  {journal} {Phys. Rev. Lett.}\ }\textbf
  {\bibinfo {volume} {128}},\ \bibinfo {pages} {195302} (\bibinfo {year}
  {2022})}\BibitemShut {NoStop}%
\bibitem [{\citenamefont {Lima}\ and\ \citenamefont
  {Pelster}(2011)}]{Lima2011a}%
  \BibitemOpen
  \bibfield  {author} {\bibinfo {author} {\bibfnamefont {A.~R.~P.}\
  \bibnamefont {Lima}}\ and\ \bibinfo {author} {\bibfnamefont {A.}~\bibnamefont
  {Pelster}},\ }\bibfield  {title} {\bibinfo {title} {Quantum fluctuations in
  dipolar {Bose} gases},\ }\href {https://doi.org/10.1103/PhysRevA.84.041604}
  {\bibfield  {journal} {\bibinfo  {journal} {Phys. Rev. A}\ }\textbf {\bibinfo
  {volume} {84}},\ \bibinfo {pages} {041604(R)} (\bibinfo {year}
  {2011})}\BibitemShut {NoStop}%
\bibitem [{\citenamefont {Lima}\ and\ \citenamefont
  {Pelster}(2012)}]{Lima2012a}%
  \BibitemOpen
  \bibfield  {author} {\bibinfo {author} {\bibfnamefont {A.~R.~P.}\
  \bibnamefont {Lima}}\ and\ \bibinfo {author} {\bibfnamefont {A.}~\bibnamefont
  {Pelster}},\ }\bibfield  {title} {\bibinfo {title} {Beyond mean-field
  low-lying excitations of dipolar {Bose} gases},\ }\href
  {https://doi.org/10.1103/PhysRevA.86.063609} {\bibfield  {journal} {\bibinfo
  {journal} {Phys. Rev. A}\ }\textbf {\bibinfo {volume} {86}},\ \bibinfo
  {pages} {063609} (\bibinfo {year} {2012})}\BibitemShut {NoStop}%
\bibitem [{\citenamefont {Ferrier-Barbut}\ \emph {et~al.}(2016)\citenamefont
  {Ferrier-Barbut}, \citenamefont {Kadau}, \citenamefont {Schmitt},
  \citenamefont {Wenzel},\ and\ \citenamefont {Pfau}}]{Ferrier-Barbut2016a}%
  \BibitemOpen
  \bibfield  {author} {\bibinfo {author} {\bibfnamefont {I.}~\bibnamefont
  {Ferrier-Barbut}}, \bibinfo {author} {\bibfnamefont {H.}~\bibnamefont
  {Kadau}}, \bibinfo {author} {\bibfnamefont {M.}~\bibnamefont {Schmitt}},
  \bibinfo {author} {\bibfnamefont {M.}~\bibnamefont {Wenzel}},\ and\ \bibinfo
  {author} {\bibfnamefont {T.}~\bibnamefont {Pfau}},\ }\bibfield  {title}
  {\bibinfo {title} {Observation of quantum droplets in a strongly dipolar
  {Bose} gas},\ }\href {https://doi.org/10.1103/PhysRevLett.116.215301}
  {\bibfield  {journal} {\bibinfo  {journal} {Phys. Rev. Lett.}\ }\textbf
  {\bibinfo {volume} {116}},\ \bibinfo {pages} {215301} (\bibinfo {year}
  {2016})}\BibitemShut {NoStop}%
\bibitem [{\citenamefont {W\"achtler}\ and\ \citenamefont
  {Santos}(2016{\natexlab{a}})}]{Wachtler2016a}%
  \BibitemOpen
  \bibfield  {author} {\bibinfo {author} {\bibfnamefont {F.}~\bibnamefont
  {W\"achtler}}\ and\ \bibinfo {author} {\bibfnamefont {L.}~\bibnamefont
  {Santos}},\ }\bibfield  {title} {\bibinfo {title} {Quantum filaments in
  dipolar {Bose}-{Einstein} condensates},\ }\href
  {https://doi.org/10.1103/PhysRevA.93.061603} {\bibfield  {journal} {\bibinfo
  {journal} {Phys. Rev. A}\ }\textbf {\bibinfo {volume} {93}},\ \bibinfo
  {pages} {061603} (\bibinfo {year} {2016}{\natexlab{a}})}\BibitemShut
  {NoStop}%
\bibitem [{\citenamefont {Bisset}\ \emph {et~al.}(2016)\citenamefont {Bisset},
  \citenamefont {Wilson}, \citenamefont {Baillie},\ and\ \citenamefont
  {Blakie}}]{Bisset2016a}%
  \BibitemOpen
  \bibfield  {author} {\bibinfo {author} {\bibfnamefont {R.~N.}\ \bibnamefont
  {Bisset}}, \bibinfo {author} {\bibfnamefont {R.~M.}\ \bibnamefont {Wilson}},
  \bibinfo {author} {\bibfnamefont {D.}~\bibnamefont {Baillie}},\ and\ \bibinfo
  {author} {\bibfnamefont {P.~B.}\ \bibnamefont {Blakie}},\ }\bibfield  {title}
  {\bibinfo {title} {Ground-state phase diagram of a dipolar condensate with
  quantum fluctuations},\ }\href {https://doi.org/10.1103/PhysRevA.94.033619}
  {\bibfield  {journal} {\bibinfo  {journal} {Phys. Rev. A}\ }\textbf {\bibinfo
  {volume} {94}},\ \bibinfo {pages} {033619} (\bibinfo {year}
  {2016})}\BibitemShut {NoStop}%
\bibitem [{\citenamefont {Saito}(2016)}]{Saito2016a}%
  \BibitemOpen
  \bibfield  {author} {\bibinfo {author} {\bibfnamefont {H.}~\bibnamefont
  {Saito}},\ }\bibfield  {title} {\bibinfo {title} {Path-integral {Monte}
  {Carlo} study on a droplet of a dipolar {Bose}-{Einstein} condensate
  stabilized by quantum fluctuation},\ }\href
  {https://doi.org/10.7566/JPSJ.85.053001} {\bibfield  {journal} {\bibinfo
  {journal} {J. Phys. Soc. Jpn}\ }\textbf {\bibinfo {volume} {85}},\ \bibinfo
  {pages} {053001} (\bibinfo {year} {2016})}\BibitemShut {NoStop}%
\bibitem [{\citenamefont {Bao}\ and\ \citenamefont {Du}(2004)}]{Bao2004a}%
  \BibitemOpen
  \bibfield  {author} {\bibinfo {author} {\bibfnamefont {W.}~\bibnamefont
  {Bao}}\ and\ \bibinfo {author} {\bibfnamefont {Q.}~\bibnamefont {Du}},\
  }\bibfield  {title} {\bibinfo {title} {Computing the ground state solution of
  {Bose}--{Einstein} condensates by a normalized gradient flow},\ }\href
  {https://doi.org/10.1137/S1064827503422956} {\bibfield  {journal} {\bibinfo
  {journal} {SIAM J. Sci. Comput.}\ }\textbf {\bibinfo {volume} {25}},\
  \bibinfo {pages} {1674} (\bibinfo {year} {2004})}\BibitemShut {NoStop}%
\bibitem [{\citenamefont {Lee}\ \emph {et~al.}(2021)\citenamefont {Lee},
  \citenamefont {Baillie},\ and\ \citenamefont {Blakie}}]{Lee2021a}%
  \BibitemOpen
  \bibfield  {author} {\bibinfo {author} {\bibfnamefont {A.-C.}\ \bibnamefont
  {Lee}}, \bibinfo {author} {\bibfnamefont {D.}~\bibnamefont {Baillie}},\ and\
  \bibinfo {author} {\bibfnamefont {P.~B.}\ \bibnamefont {Blakie}},\ }\bibfield
   {title} {\bibinfo {title} {Numerical calculation of dipolar-quantum-droplet
  stationary states},\ }\href
  {https://doi.org/10.1103/PhysRevResearch.3.013283} {\bibfield  {journal}
  {\bibinfo  {journal} {Phys. Rev. Research}\ }\textbf {\bibinfo {volume}
  {3}},\ \bibinfo {pages} {013283} (\bibinfo {year} {2021})}\BibitemShut
  {NoStop}%
\bibitem [{\citenamefont {Leggett}(1970)}]{Leggett1970a}%
  \BibitemOpen
  \bibfield  {author} {\bibinfo {author} {\bibfnamefont {A.~J.}\ \bibnamefont
  {Leggett}},\ }\bibfield  {title} {\bibinfo {title} {Can a solid be
  `superfluid'?},\ }\href {https://doi.org/10.1103/PhysRevLett.25.1543}
  {\bibfield  {journal} {\bibinfo  {journal} {Phys. Rev. Lett.}\ }\textbf
  {\bibinfo {volume} {25}},\ \bibinfo {pages} {1543} (\bibinfo {year}
  {1970})}\BibitemShut {NoStop}%
\bibitem [{\citenamefont {Leggett}(1998)}]{Leggett1998a}%
  \BibitemOpen
  \bibfield  {author} {\bibinfo {author} {\bibfnamefont {A.~J.}\ \bibnamefont
  {Leggett}},\ }\bibfield  {title} {\bibinfo {title} {On the superfluid
  fraction of an arbitrary many-body system at {$T=0$}},\ }\href
  {https://doi.org/10.1023/B:JOSS.0000033170.38619.6c} {\bibfield  {journal}
  {\bibinfo  {journal} {J. Stat. Phys.}\ }\textbf {\bibinfo {volume} {93}},\
  \bibinfo {pages} {927} (\bibinfo {year} {1998})}\BibitemShut {NoStop}%
\bibitem [{\citenamefont {Ancilotto}\ \emph {et~al.}(2013)\citenamefont
  {Ancilotto}, \citenamefont {Rossi},\ and\ \citenamefont
  {Toigo}}]{Ancilotto2013a}%
  \BibitemOpen
  \bibfield  {author} {\bibinfo {author} {\bibfnamefont {F.}~\bibnamefont
  {Ancilotto}}, \bibinfo {author} {\bibfnamefont {M.}~\bibnamefont {Rossi}},\
  and\ \bibinfo {author} {\bibfnamefont {F.}~\bibnamefont {Toigo}},\ }\bibfield
   {title} {\bibinfo {title} {Supersolid structure and excitation spectrum of
  soft-core bosons in three dimensions},\ }\href
  {https://doi.org/10.1103/PhysRevA.88.033618} {\bibfield  {journal} {\bibinfo
  {journal} {Phys. Rev. A}\ }\textbf {\bibinfo {volume} {88}},\ \bibinfo
  {pages} {033618} (\bibinfo {year} {2013})}\BibitemShut {NoStop}%
\bibitem [{\citenamefont {Saslow}(1976)}]{Saslow1976a}%
  \BibitemOpen
  \bibfield  {author} {\bibinfo {author} {\bibfnamefont {W.~M.}\ \bibnamefont
  {Saslow}},\ }\bibfield  {title} {\bibinfo {title} {Superfluidity of periodic
  solids},\ }\href {https://doi.org/10.1103/PhysRevLett.36.1151} {\bibfield
  {journal} {\bibinfo  {journal} {Phys. Rev. Lett.}\ }\textbf {\bibinfo
  {volume} {36}},\ \bibinfo {pages} {1151} (\bibinfo {year}
  {1976})}\BibitemShut {NoStop}%
\bibitem [{\citenamefont {Blakie}()}]{Blakie2023b}%
  \BibitemOpen
  \bibfield  {author} {\bibinfo {author} {\bibfnamefont {P.~B.}\ \bibnamefont
  {Blakie}},\ }\href@noop {} {\bibinfo {title} {Superfluid fraction tensor of a
  two-dimensional supersolid}},\ \Eprint {https://arxiv.org/abs/2308.14001}
  {arXiv:2308.14001} \BibitemShut {NoStop}%
\bibitem [{\citenamefont {Sep{\'u}lveda}\ \emph {et~al.}(2010)\citenamefont
  {Sep{\'u}lveda}, \citenamefont {Josserand},\ and\ \citenamefont
  {Rica}}]{Sepulveda2010a}%
  \BibitemOpen
  \bibfield  {author} {\bibinfo {author} {\bibfnamefont {N.}~\bibnamefont
  {Sep{\'u}lveda}}, \bibinfo {author} {\bibfnamefont {C.}~\bibnamefont
  {Josserand}},\ and\ \bibinfo {author} {\bibfnamefont {S.}~\bibnamefont
  {Rica}},\ }\bibfield  {title} {\bibinfo {title} {Superfluid density in a
  two-dimensional model of supersolid},\ }\href
  {https://doi.org/10.1140/epjb/e2010-10176-y} {\bibfield  {journal} {\bibinfo
  {journal} {EPJ B}\ }\textbf {\bibinfo {volume} {78}},\ \bibinfo {pages} {439}
  (\bibinfo {year} {2010})}\BibitemShut {NoStop}%
\bibitem [{\citenamefont {Baillie}\ \emph {et~al.}(2017)\citenamefont
  {Baillie}, \citenamefont {Wilson},\ and\ \citenamefont
  {Blakie}}]{Baillie2017a}%
  \BibitemOpen
  \bibfield  {author} {\bibinfo {author} {\bibfnamefont {D.}~\bibnamefont
  {Baillie}}, \bibinfo {author} {\bibfnamefont {R.~M.}\ \bibnamefont
  {Wilson}},\ and\ \bibinfo {author} {\bibfnamefont {P.~B.}\ \bibnamefont
  {Blakie}},\ }\bibfield  {title} {\bibinfo {title} {Collective excitations of
  self-bound droplets of a dipolar quantum fluid},\ }\href
  {https://doi.org/10.1103/PhysRevLett.119.255302} {\bibfield  {journal}
  {\bibinfo  {journal} {Phys. Rev. Lett.}\ }\textbf {\bibinfo {volume} {119}},\
  \bibinfo {pages} {255302} (\bibinfo {year} {2017})}\BibitemShut {NoStop}%
\bibitem [{\citenamefont {Baillie}\ and\ \citenamefont
  {Blakie}(2015)}]{Baillie2015b}%
  \BibitemOpen
  \bibfield  {author} {\bibinfo {author} {\bibfnamefont {D.}~\bibnamefont
  {Baillie}}\ and\ \bibinfo {author} {\bibfnamefont {P.~B.}\ \bibnamefont
  {Blakie}},\ }\bibfield  {title} {\bibinfo {title} {A general theory of
  flattened dipolar condensates},\ }\href
  {https://doi.org/10.1088/1367-2630/17/3/033028} {\bibfield  {journal}
  {\bibinfo  {journal} {New J. Phys.}\ }\textbf {\bibinfo {volume} {17}},\
  \bibinfo {pages} {033028} (\bibinfo {year} {2015})}\BibitemShut {NoStop}%
\bibitem [{\citenamefont {Prestipino}\ \emph
  {et~al.}(2018{\natexlab{b}})\citenamefont {Prestipino}, \citenamefont
  {Sergi},\ and\ \citenamefont {Bruno}}]{Prestipino2019a}%
  \BibitemOpen
  \bibfield  {author} {\bibinfo {author} {\bibfnamefont {S.}~\bibnamefont
  {Prestipino}}, \bibinfo {author} {\bibfnamefont {A.}~\bibnamefont {Sergi}},\
  and\ \bibinfo {author} {\bibfnamefont {E.}~\bibnamefont {Bruno}},\ }\bibfield
   {title} {\bibinfo {title} {Clusterization of weakly-interacting bosons in
  one dimension: an analytic study at zero temperature},\ }\href
  {https://doi.org/10.1088/1751-8121/aaee94} {\bibfield  {journal} {\bibinfo
  {journal} {J. Phys. A}\ }\textbf {\bibinfo {volume} {52}},\ \bibinfo {pages}
  {015002} (\bibinfo {year} {2018}{\natexlab{b}})}\BibitemShut {NoStop}%
\bibitem [{\citenamefont {Lima}\ and\ \citenamefont
  {Pelster}(2010)}]{Lima2010a}%
  \BibitemOpen
  \bibfield  {author} {\bibinfo {author} {\bibfnamefont {A.~R.~P.}\
  \bibnamefont {Lima}}\ and\ \bibinfo {author} {\bibfnamefont {A.}~\bibnamefont
  {Pelster}},\ }\bibfield  {title} {\bibinfo {title} {Dipolar {Fermi} gases in
  anisotropic traps},\ }\href {https://doi.org/10.1103/PhysRevA.81.063629}
  {\bibfield  {journal} {\bibinfo  {journal} {Phys. Rev. A}\ }\textbf {\bibinfo
  {volume} {81}},\ \bibinfo {pages} {063629} (\bibinfo {year}
  {2010})}\BibitemShut {NoStop}%
\bibitem [{\citenamefont {Ticknor}\ \emph {et~al.}(2011)\citenamefont
  {Ticknor}, \citenamefont {Wilson},\ and\ \citenamefont
  {Bohn}}]{Ticknor2011a}%
  \BibitemOpen
  \bibfield  {author} {\bibinfo {author} {\bibfnamefont {C.}~\bibnamefont
  {Ticknor}}, \bibinfo {author} {\bibfnamefont {R.~M.}\ \bibnamefont
  {Wilson}},\ and\ \bibinfo {author} {\bibfnamefont {J.~L.}\ \bibnamefont
  {Bohn}},\ }\bibfield  {title} {\bibinfo {title} {Anisotropic superfluidity in
  a dipolar {Bose} gas},\ }\href
  {https://doi.org/10.1103/PhysRevLett.106.065301} {\bibfield  {journal}
  {\bibinfo  {journal} {Phys. Rev. Lett.}\ }\textbf {\bibinfo {volume} {106}},\
  \bibinfo {pages} {065301} (\bibinfo {year} {2011})}\BibitemShut {NoStop}%
\bibitem [{\citenamefont {Wenzel}\ \emph {et~al.}(2018)\citenamefont {Wenzel},
  \citenamefont {B\"ottcher}, \citenamefont {Schmidt}, \citenamefont
  {Eisenmann}, \citenamefont {Langen}, \citenamefont {Pfau},\ and\
  \citenamefont {Ferrier-Barbut}}]{Wenzel2018a}%
  \BibitemOpen
  \bibfield  {author} {\bibinfo {author} {\bibfnamefont {M.}~\bibnamefont
  {Wenzel}}, \bibinfo {author} {\bibfnamefont {F.}~\bibnamefont {B\"ottcher}},
  \bibinfo {author} {\bibfnamefont {J.-N.}\ \bibnamefont {Schmidt}}, \bibinfo
  {author} {\bibfnamefont {M.}~\bibnamefont {Eisenmann}}, \bibinfo {author}
  {\bibfnamefont {T.}~\bibnamefont {Langen}}, \bibinfo {author} {\bibfnamefont
  {T.}~\bibnamefont {Pfau}},\ and\ \bibinfo {author} {\bibfnamefont
  {I.}~\bibnamefont {Ferrier-Barbut}},\ }\bibfield  {title} {\bibinfo {title}
  {Anisotropic superfluid behavior of a dipolar {Bose}-{Einstein} condensate},\
  }\href {https://doi.org/10.1103/PhysRevLett.121.030401} {\bibfield  {journal}
  {\bibinfo  {journal} {Phys. Rev. Lett.}\ }\textbf {\bibinfo {volume} {121}},\
  \bibinfo {pages} {030401} (\bibinfo {year} {2018})}\BibitemShut {NoStop}%
\bibitem [{\citenamefont {Bombin}\ \emph {et~al.}(2017)\citenamefont {Bombin},
  \citenamefont {Boronat},\ and\ \citenamefont {Mazzanti}}]{Bombin2017a}%
  \BibitemOpen
  \bibfield  {author} {\bibinfo {author} {\bibfnamefont {R.}~\bibnamefont
  {Bombin}}, \bibinfo {author} {\bibfnamefont {J.}~\bibnamefont {Boronat}},\
  and\ \bibinfo {author} {\bibfnamefont {F.}~\bibnamefont {Mazzanti}},\
  }\bibfield  {title} {\bibinfo {title} {Dipolar {Bose} supersolid stripes},\
  }\href {https://doi.org/10.1103/PhysRevLett.119.250402} {\bibfield  {journal}
  {\bibinfo  {journal} {Phys. Rev. Lett.}\ }\textbf {\bibinfo {volume} {119}},\
  \bibinfo {pages} {250402} (\bibinfo {year} {2017})}\BibitemShut {NoStop}%
\bibitem [{\citenamefont {Cinti}\ and\ \citenamefont
  {Boninsegni}(2019)}]{Cinti2019a}%
  \BibitemOpen
  \bibfield  {author} {\bibinfo {author} {\bibfnamefont {F.}~\bibnamefont
  {Cinti}}\ and\ \bibinfo {author} {\bibfnamefont {M.}~\bibnamefont
  {Boninsegni}},\ }\bibfield  {title} {\bibinfo {title} {Absence of
  superfluidity in 2{D} dipolar {B}ose striped crystals},\ }\href
  {https://doi.org/10.1007/s10909-019-02209-3} {\bibfield  {journal} {\bibinfo
  {journal} {J. Low Temp. Phys.}\ }\textbf {\bibinfo {volume} {196}},\ \bibinfo
  {pages} {413} (\bibinfo {year} {2019})}\BibitemShut {NoStop}%
\bibitem [{\citenamefont {Cinti}\ and\ \citenamefont
  {Boninsegni}(2020)}]{Cinti2020a}%
  \BibitemOpen
  \bibfield  {author} {\bibinfo {author} {\bibfnamefont {F.}~\bibnamefont
  {Cinti}}\ and\ \bibinfo {author} {\bibfnamefont {M.}~\bibnamefont
  {Boninsegni}},\ }\bibfield  {title} {\bibinfo {title} {Comment on
  ``{Berezinskii}-{Kosterlitz}-{Thouless} transition in two-dimensional dipolar
  stripes''},\ }\href {https://doi.org/10.1103/PhysRevA.102.047301} {\bibfield
  {journal} {\bibinfo  {journal} {Phys. Rev. A}\ }\textbf {\bibinfo {volume}
  {102}},\ \bibinfo {pages} {047301} (\bibinfo {year} {2020})}\BibitemShut
  {NoStop}%
\bibitem [{\citenamefont {Bomb\'{\i}n}\ \emph {et~al.}(2020)\citenamefont
  {Bomb\'{\i}n}, \citenamefont {Mazzanti},\ and\ \citenamefont
  {Boronat}}]{Bombin2020a}%
  \BibitemOpen
  \bibfield  {author} {\bibinfo {author} {\bibfnamefont {R.}~\bibnamefont
  {Bomb\'{\i}n}}, \bibinfo {author} {\bibfnamefont {F.}~\bibnamefont
  {Mazzanti}},\ and\ \bibinfo {author} {\bibfnamefont {J.}~\bibnamefont
  {Boronat}},\ }\bibfield  {title} {\bibinfo {title} {Reply to ``comment on
  `{Berezinskii}-{Kosterlitz}-{Thouless} transition in two-dimensional dipolar
  stripes' ''},\ }\href {https://doi.org/10.1103/PhysRevA.102.047302}
  {\bibfield  {journal} {\bibinfo  {journal} {Phys. Rev. A}\ }\textbf {\bibinfo
  {volume} {102}},\ \bibinfo {pages} {047302} (\bibinfo {year}
  {2020})}\BibitemShut {NoStop}%
\bibitem [{\citenamefont {Chauveau}\ \emph {et~al.}(2023)\citenamefont
  {Chauveau}, \citenamefont {Maury}, \citenamefont {Rabec}, \citenamefont
  {Heintze}, \citenamefont {Brochier}, \citenamefont {Nascimbene},
  \citenamefont {Dalibard}, \citenamefont {Beugnon}, \citenamefont {Roccuzzo},\
  and\ \citenamefont {Stringari}}]{Chauveau2023a}%
  \BibitemOpen
  \bibfield  {author} {\bibinfo {author} {\bibfnamefont {G.}~\bibnamefont
  {Chauveau}}, \bibinfo {author} {\bibfnamefont {C.}~\bibnamefont {Maury}},
  \bibinfo {author} {\bibfnamefont {F.}~\bibnamefont {Rabec}}, \bibinfo
  {author} {\bibfnamefont {C.}~\bibnamefont {Heintze}}, \bibinfo {author}
  {\bibfnamefont {G.}~\bibnamefont {Brochier}}, \bibinfo {author}
  {\bibfnamefont {S.}~\bibnamefont {Nascimbene}}, \bibinfo {author}
  {\bibfnamefont {J.}~\bibnamefont {Dalibard}}, \bibinfo {author}
  {\bibfnamefont {J.}~\bibnamefont {Beugnon}}, \bibinfo {author} {\bibfnamefont
  {S.~M.}\ \bibnamefont {Roccuzzo}},\ and\ \bibinfo {author} {\bibfnamefont
  {S.}~\bibnamefont {Stringari}},\ }\bibfield  {title} {\bibinfo {title}
  {Superfluid fraction in an interacting spatially modulated {Bose}-{Einstein}
  condensate},\ }\href {https://doi.org/10.1103/PhysRevLett.130.226003}
  {\bibfield  {journal} {\bibinfo  {journal} {Phys. Rev. Lett.}\ }\textbf
  {\bibinfo {volume} {130}},\ \bibinfo {pages} {226003} (\bibinfo {year}
  {2023})}\BibitemShut {NoStop}%
\bibitem [{\citenamefont {Tao}\ \emph {et~al.}(2023)\citenamefont {Tao},
  \citenamefont {Zhao},\ and\ \citenamefont {Spielman}}]{Tao2023a}%
  \BibitemOpen
  \bibfield  {author} {\bibinfo {author} {\bibfnamefont {J.}~\bibnamefont
  {Tao}}, \bibinfo {author} {\bibfnamefont {M.}~\bibnamefont {Zhao}},\ and\
  \bibinfo {author} {\bibfnamefont {I.~B.}\ \bibnamefont {Spielman}},\
  }\bibfield  {title} {\bibinfo {title} {Observation of anisotropic superfluid
  density in an artificial crystal},\ }\href
  {https://doi.org/10.1103/PhysRevLett.131.163401} {\bibfield  {journal}
  {\bibinfo  {journal} {Phys. Rev. Lett.}\ }\textbf {\bibinfo {volume} {131}},\
  \bibinfo {pages} {163401} (\bibinfo {year} {2023})}\BibitemShut {NoStop}%
\bibitem [{\citenamefont {W\"achtler}\ and\ \citenamefont
  {Santos}(2016{\natexlab{b}})}]{Wachtler2016b}%
  \BibitemOpen
  \bibfield  {author} {\bibinfo {author} {\bibfnamefont {F.}~\bibnamefont
  {W\"achtler}}\ and\ \bibinfo {author} {\bibfnamefont {L.}~\bibnamefont
  {Santos}},\ }\bibfield  {title} {\bibinfo {title} {Ground-state properties
  and elementary excitations of quantum droplets in dipolar {Bose}-{Einstein}
  condensates},\ }\href {https://doi.org/10.1103/PhysRevA.94.043618} {\bibfield
   {journal} {\bibinfo  {journal} {Phys. Rev. A}\ }\textbf {\bibinfo {volume}
  {94}},\ \bibinfo {pages} {043618} (\bibinfo {year}
  {2016}{\natexlab{b}})}\BibitemShut {NoStop}%
\bibitem [{\citenamefont {Zhang}\ and\ \citenamefont
  {Maucher}(2023)}]{Zhang2023a}%
  \BibitemOpen
  \bibfield  {author} {\bibinfo {author} {\bibfnamefont {Y.-C.}\ \bibnamefont
  {Zhang}}\ and\ \bibinfo {author} {\bibfnamefont {F.}~\bibnamefont
  {Maucher}},\ }\bibfield  {title} {\bibinfo {title} {Variational approaches to
  two-dimensionally symmetry-broken dipolar {Bose}-{Einstein} condensates},\
  }\href {https://doi.org/10.3390/atoms11070102} {\bibfield  {journal}
  {\bibinfo  {journal} {Atoms}\ }\textbf {\bibinfo {volume} {11}},\ \bibinfo
  {pages} {102} (\bibinfo {year} {2023})}\BibitemShut {NoStop}%
\bibitem [{\citenamefont {Zhang}\ \emph {et~al.}()\citenamefont {Zhang},
  \citenamefont {Pohl},\ and\ \citenamefont {Maucher}}]{Zhang2023b}%
  \BibitemOpen
  \bibfield  {author} {\bibinfo {author} {\bibfnamefont {Y.-C.}\ \bibnamefont
  {Zhang}}, \bibinfo {author} {\bibfnamefont {T.}~\bibnamefont {Pohl}},\ and\
  \bibinfo {author} {\bibfnamefont {F.}~\bibnamefont {Maucher}},\ }\bibfield
  {title} {\bibinfo {title} {Metastable patterns in one- and two-component
  dipolar {Bose}-{Einstein} condensates},\ }\href@noop {} {\ }\Eprint
  {https://arxiv.org/abs/2310.04738} {arXiv:2310.04738} \BibitemShut {NoStop}%
\end{thebibliography}


\end{document}